\documentclass[a4paper,fleqn,usenatbib]{mnras}

\usepackage{newtxtext,newtxmath}

\usepackage[T1]{fontenc}
\usepackage{ae,aecompl}
\usepackage{relsize} 
\usepackage{color,soul}
\usepackage{subfig}

\usepackage{xcolor}

\usepackage{graphicx}
\usepackage{amsmath}
\usepackage{bm}
\usepackage{footnote}
\usepackage{bigints}
\usepackage{colortbl}

\definecolor{Gray}{gray}{0.9}
\defcitealias{Gie11}{GH11}
\defcitealias{Gil00}{GBG+00}
\defcitealias{Hamers17}{HT17}
\defcitealias{PaperI}{Paper I}

\title[Multiple populations via sub-stellar accretion]{\vspace{-2mm}Accretion of sub-stellar companions as the origin of chemical abundance inhomogeneities in globular clusters\vspace{-3mm}}

\author[Winter \& Clarke]{Andrew J. Winter\thanks{andrew.winter@oca.eu}$^{1,2}$ and Cathie J. Clarke$^{3}$\\
$^{1}$Universit\'{e} C\^{o}te d'Azur, Observatoire de la C\^{o}te d'Azur, CNRS, Laboratoire Lagrange, 06300 Nice, France\\
$^{2}$Universit\'{e} Grenoble Alpes, CNRS, IPAG, 38000 Grenoble, France\\
$^{3}$Institute of Astronomy, University of Cambridge, Madingley Road, Cambridge, CB3 0HA, UK }\vspace{-3mm}

\date{Accepted 2023 January 25. Received 2023 January 08; in original form 2022 August 05}\vspace{-2mm}

\pubyear{2021}

\begin{document}
\label{firstpage}
\pagerange{\pageref{firstpage}--\pageref{lastpage}}
\maketitle

\begin{abstract}
{Globular clusters exhibit abundance variations, defining `multiple populations', which have prompted a protracted search for their origin. Properties requiring explanation include: the high fraction of polluted stars ($\sim 40{-}90$~percent, correlated with cluster mass), the absence of pollution in young clusters and the lower pollution rate with binarity and distance from the cluster centre. We present a novel mechanism for late delivery of pollutants into stars via accretion of sub-stellar companions. In this scenario, stars move through a medium polluted with AGB and massive star ejecta, accreting material to produce companions with typical mass ratio $q\sim 0.1$. These companions undergo eccentricity excitation due to dynamical perturbations by passing stars, culminating in a merger with their host star. The accretion of the companion alters surface abundances via injected pollutant. {Alongside other self-enrichment models, the companion accretion model can explain the dilution of pollutant and correlation with intra-cluster location. The model also explains the ubiquity and discreteness of the populations and correlations of enrichment rates with cluster mass, cluster age and stellar binarity. Abundance variations in some clusters can be broadly reproduced using AGB and massive binary ejecta abundances from the literature. In other clusters, some high companion mass ratios ($q\gtrsim 1$) are required. In these cases, the available mass budget necessitates a variable degree of mixing of the polluted material with the primary star, deviations from model ejecta abundances or mixing of internal burning products.} We highlight the avenues of further investigation which are required to explore some of the key processes invoked in this model.}
\end{abstract}

\begin{keywords} 
globular clusters: general -- binaries: general -- stars: abundances -- planets and satellites: formation,  dynamical evolution and stability\vspace{-2mm}
\end{keywords}


\section{Introduction}

Ubiquitous, discretised and complex structure in the abundance distributions of stars in old, massive globular clusters have remained a challenging puzzle for many decades \citep[e.g.][]{Cohen78, Dickens79, Kraft94, Gratton04, Bastian18}. While the specific abundance variations vary significantly from cluster to cluster, a common trend in the abundance variations relative to the primordial stellar population is N enhancement, which combines with O and C depletion to yield approximately constant C+N+O abundances \citep[e.g.][]{Dickens91}. These variations are also correlated or anti-correlated with He, Na, Al and Mg abundances \citep[see review by][]{Gratton04}. The occurrence of these discretised variations is correlated with numerous cluster and stellar properties, including (but not limited to) the cluster mass \citep[e.g.][]{Milone17} and age \citep{Martocchia18a}, and the binarity of a given star \citep[e.g.][]{DOrazi15}.

Despite numerous mechanisms put forward for the origin of multiple populations in globular clusters, to date all fail to explain numerous properties without violating other empirical constraints \citep[for a review, see][]{Bastian18}. One of the most stringent of these constraints is known as the mass budget problem (see Section~\ref{sec:enrfrac_mbudget}), which is that the polluted population comprises up to $f_\mathrm{II} \sim 90$~percent of the total mass of the present day cluster. Many explanations for abundance variation hinge on the formation of a secondary population from the ejecta of massive or asymptotic giant branch (AGB) stars. However, the maximum mass available from these stars to form this second population is $\lesssim 10$~percent of the accompanying mass of low mass stars, long-lived stars. This leads to a challenging problem in producing the observed population, which cannot be explained by many current models.

In this work, we propose an origin for the abundance variations in globular clusters that helps solve this mass budget problem, as well as explaining the origin for many of the empirical correlations associated with multiple populations. This mechanism takes inspiration from recent results suggesting that sub-stellar companions to pulsating red giant stars are the origin of the long secondary period variability \citep[][see also \citealt{Beck14}]{Soszynski21}. Assuming this explains all such variability, such sub-stellar companions would also seem to exist in globular clusters \citep{Percy21}. Here we make use of our recently developed analytic expressions describing how such companions evolve dynamically in dense environments \citep{Winter22c}, in order to connect these apparently unrelated phenomena: sub-stellar companions and multiple stellar populations. 

In brief, our proposed scenario proceeds in the following way. Sub-stellar companions with mass ratio $q\sim 0.1$ form from AGB or massive star ejecta {via tidal capture, disc sweep-up or Bondi-Hoyle-Lyttleton accretion.} They are then subject to extreme eccentricity excitation resulting from many distant encounters in the dense core of the stellar cluster. Eventually, this eccentricity evolution leads to a collision with the host star. This collision induces deep (rotational) mixing, allowing pollution from the companion (and possibly internal fusion products) to produce surface abundance variations in the primary. 

In the following manuscript, we explore this companion accretion model quantitatively and qualitatively in terms of the observational constraints. In doing so, we demonstrate both the feasibility of the model and the advantages over many existing models. In Section~\ref{sec:obs_req} we discuss the observational requirements of any successful model for producing multiple populations in more detail. We explain the predictions and requirements of our model in Section~\ref{sec:model}, making comparisons with the existing observational constraints. We draw our conclusions in Section~\ref{sec:concs}, outlining possible future directions that may further test the companion accretion model.

\section{Observational requirements}
\label{sec:obs_req}
\subsection{Discrete elemental abundance variations}

The defining feature of the multiple populations found in globular clusters is their elemental abundance variations. These variations are nearly ubiquitous in all massive and old clusters. In general, the polluted stars are characterised by enhancement of He, N and Na and depletion of O and C, although the specific variations are unique to each globular cluster. We do not review the specific variations in detail here, but refer the reader to the review by \citet{Bastian18}.  However, we highlight that abundance spreads are limited to light elements, with little star-to-star Fe or heavy element variation. This suggests a unique chemical processing mechanism applies only in dense cluster environments. 

While the abundance variations in each cluster are unique, some prevailing correlations and anti-correlations appear to be common. Enhancements in N are associated with depletion in C and O, such that the overall C+N+O abundance remains approximately constant \citep[e.g.][]{Dickens91}. Meanwhile, enhanced N abundance (depleted C, O) is positively correlated with Na \cite[e.g.][]{Sneden92}. A weak anti-correlation may also be apparent between enhanced Al and depleted Mg  \citep[see review by][]{Gratton04}. Changes in the total He abundance $\Delta Y$ are most reliably inferred from MS isochrone fitting \citep{Cassisi17}. Enrichment of He exhibits a positive correlation with cluster mass, and typical spreads in abundance $\Delta Y \sim 0.1{-}0.2$ from the pristine $Y=0.24$ \citep[e.g.][]{King12, Milone15}. Finally and significantly, Li depletion that is expected from the hot H burning that produces N, Na and Al is found in some instances \citep{DOrazi15}, but appears not to be universal \citep[e.g.][]{Mucciarelli11}. {In clusters for which Li abundances are approximately constant through the MS, this would suggest that a large quantity of pristine material to dilute the pollutant is necessary.}

An interesting property of the chemical enrichment is its apparent {bimodality (or multi-modality)}. Multi-modality in the abundance distribution is particularly seen in the C and N abundances, for which sub-populations are apparent almost ubiquitously \citep[where errors on CN measurements are small enough -- e.g.][]{Norris87}. Discrete sub-divisions can also be seen in high resolution observations constraining abundances of O, Na and Al \citep[e.g.][]{Marino08, Lind11}. Thus, any pollution must proceed in a discrete way, which may be a problem for many proposed scenarios, such as early disc accretion \citep{Bastian13}.

Another important property of the variations in the chemical abundances is that pollutants are mixed through the majority of the star. This can be inferred from the near constant abundances observed through the main sequence (MS), main sequence turn-off (MSTO) and red giant branch (RGB) phases \citep[e.g.][]{Briley04}. Such stars all have radically different convective zones,  {varying from $\sim 1$~percent of the mass for MSTO stars to $\sim 70$~percent of the mass for RGB stars} \citep[see Figure 1 of][]{Briley04}. Thus, surface pollution is ruled out as the origin, and this has lead to the conclusion for many authors that the stars must form directly out of the polluted material as part of a second generation. Assuming that each of the populations represent different generations comes with its own problems, including the pervasive mass budget problem (Section~\ref{sec:enrfrac_mbudget}).

\subsection{Enrichment fraction and cluster properties}

The typical  fraction of chemically enriched stars in globular clusters is $f_\mathrm{II} \sim 0.4{-}0.9$, and is most robustly positively correlated with the total mass of the cluster \citep[e.g.][]{Milone17}. Meanwhile, the polluted star fraction $f_\mathrm{II}$ is neither strongly correlated with galactocentric distance nor metallicity \citep{Bastian15a}. The former strongly suggests that enrichment comes from an internal source, while the latter suggests that the fraction is not dependent on stellar evolution but rather on some dynamical process. 

Particularly problematic for most models, the presence of multiple populations in clusters is also associated with age. The youngest globular cluster found to contain an enriched population is the $1.9\pm0.1$~Gyr old NGC 1978 \citep{Mucciarelli07, Martocchia18a, Saracino20}. By contrast, the globular cluster NGC 419 contains no evidence of an enriched population. The latter has stellar mass $M_\mathrm{c} \approx  8\cdot 10^4 \, M_\odot$ \citep{Song21}, half-mass radius $R_\mathrm{hm} \approx 8.1$~pc \citep{Glatt09} and an age of $t=1.5$~Gyr \citep{Glatt08}. This is close to the $M_\mathrm{c} \approx 2\times 10^5 \, M_\odot$ and $R_\mathrm{hm} = 8.7$~pc of NGC 1978 \citep[][and references therein]{Milone17}.  Meanwhile, the age spread between populations {within individual clusters} is small \citep[$<20$~Myr,][]{Martocchia18b}, indicating that the multiple populations initially form nearly simultaneously. {Combined with the absence of multiple populations in young clusters, this provides a difficult challenge for enrichment models. {Since few young clusters are very massive and dense,} one possible solution is that the absence of polluted stars at young ages is in fact related to a strict mass/radius requirement. In this instance, given the properties of the clusters in which multiple populations have not been found, then enrichment can only occur at all when $M_\mathrm{c}> 10^5\, M_\odot$ or $M_\mathrm{c}/R_\mathrm{c} > 10^4 \, M_\odot$~pc$^{-1}$. {However, multiple populations in older clusters with lower masses and larger radii than these thresholds have been found, such that one needs to appeal to much more massive initial conditions for these clusters.} Alternatively, the above findings may be reconciled if the pristine population becomes the polluted population over Gyr timescales.} 

\subsection{Mass budget problem}

\label{sec:enrfrac_mbudget}
Another severe constraint on any mechanism proposed to produce multiple populations is that it produces enough mass to make the second population. Since the mass of the second constitutes up to $\sim 90$~percent of the total stellar population, this is a particularly difficult requirement for self-enrichment models invoking a standard initial mass function \citep[e.g.][]{Prantzos06}. The total mass of AGB stars is $\sim 8$~percent of the initial stellar mass, while the mass in the low mass stars ($m_*<0.8 \, M_\odot$) is $\sim 40$~percent. Hence, if all {of the AGB star mass (an extreme assumption)} went into forming a second generation with a similar initial mass function IMF, then $93$~percent of the total mass now should be primordial (or $83$~percent if all second generation stars are low mass), while $8$~percent ($17$~percent) of the low mass IMF would have undergone processing. Even for globular clusters with a relatively large present day fraction of pristine stars $f_\mathrm{I} \sim 1/3$, we  would therefore need to increase the number of second generation stars by a factor $\sim 10$ or decrease the number of first generation stars by $95$~percent. {Alternatively, a hard minimum of a $90$~percent reduction in mass of the first generation stars corresponds to the very extreme assumption that $100$~percent of the mass within AGB stars goes into low mass second generation stars.}

Some mitigation in this regard may be expected from the fact that dilution is needed for the observed chemical abundances \citep[e.g.][]{DErcole10}. {To avoid too much iron from supernovae in the material forming the second generation, models have invoked the removal of the primordial gas by these supernovae \citep[although long tails in Fe content are observed in some globular clusters --][]{Carretta10}.} However, then the diluting material must be re-accreted onto the globular cluster from the surrounding environment after $20{-}30$~Myr \citep{DErcole16}. This secondary accretion process is expected to be inefficient and requires a large reservoir of remaining gas \citep{Conroy11}. Even in this case, the time difference between the two populations would be in tension with the apparently small age spreads between multiple populations \citep{Martocchia18b}. The fresh material accreted from the surrounding galaxy also needs to match the abundances in the pristine medium; it is unclear that this would be consistent with the observed Fe spreads \citep[e.g.][]{Bailin22}. Finally, and most importantly, the quantity of pristine gas must not greatly exceed the pollutant in order to produce the observed abundance variations \citep{DErcole11}, so this dilution process in isolation does not solve the mass budget problem. 

We might also consider the ejection of first generation stars over the life-time of the cluster. The required depletion factor is $f_\mathrm{dep} = M_\mathrm{c}/M_\mathrm{c,0} \lesssim 0.1$ (more than $90$~percent mass-loss). {This might be increased to $f_\mathrm{dep} \approx 0.25$ if we take the very extreme assumptions stated above ($100$~percent AGB stars to low mass second generation stars).} This assumes that practically all of the second generation are retained \citep{Conroy12, CabreraZiri15}. Instead, $f_\mathrm{dep}\sim 0.5$ for typical present day globular clusters is expected theoretically \citep{Kruijssen15} and from their current demographics \citep{Webb15}. A tidally disrupted cluster is required in order to reach such small $f_\mathrm{dep}$ \citep{Vesperini10}. {Even then, \citet{Larsen12} showed that this is also inconsistent with the globular cluster stars in the Fornax Dwarf galaxy, which make up $20{-}25$~percent of the total low metallicity stars including those in the field. This means that even if all of the field stars formed in globular clusters, these clusters could only have been a factor $\lesssim 4{-}5$ more massive at their birth.} Further, even if this rapid depletion could be responsible, it seems implausible that the degree of depletion would increase with increasing cluster mass. {If anything we would expect the opposite trend given that the dispersal timescale for clusters evolving in a tidal field increases with cluster mass \citep[e.g.][]{Lamers05}.} Therefore dynamical depletion alone is not sufficient to resolve the mass budget problem.

\subsection{Correlations with binarity and intra-cluster location}

The pristine and enriched stellar populations often appear not to have the same spatial distribution within their host cluster. Most studies indicate a greater degree of central concentration in the enriched population than the pristine one \citep[e.g.][]{Bellini09, Lardo11, Simioni16}, although in some cases they appear to be well-mixed \citep[e.g.][]{Dalessandro14,Vanderbeke15, Miholics15} or even less concentrated than the pristine population \citep[e.g.][]{Larsen15}.  The pristine population exhibits an enhanced binary fraction with respect to the polluted population \citep{DOrazi10,Lucatello15}.

\section{Model}
\label{sec:model}
\subsection{Overall picture}

In this section we present a novel model to explain the observed chemical enrichment for populations of stars in globular clusters. Our model is in some sense a combination of the `early disc accretion' scenario, put forward by \citet{Bastian13}, and the second generation models \citep[e.g.][]{Decressin07, DErcole08}. In the early disc accretion scenario, mass is added to the accretion disc as the young star-disc system moves through a the interstellar medium that is enriched by massive rapidly rotating stars and/or binaries. This polluted material is then accreted onto the host star, leading to inhomogeneous chemistry in the stellar population. However, this scenario suffers a number of issues, perhaps most prominently that the accretion must occur extremely rapidly while the star is fully convective to explain abundance variations that remain constant across MS, MSTO and RGB \citep[e.g.][]{Briley04}.

The scenario we explore in this work is the `companion accretion model'. Instead of requiring that all the mass accreted onto the disc in the very early stages of cluster evolution, this mechanism invokes a later delivery of polluted material. The chronology of this scenario can be summarised as follows:
\begin{enumerate}
    \item The first generation of stars forms, while material from the already formed AGB stars and massive binaries continues to be ejected. This material remains bound in the core of the cluster due to their low velocities \citep[$\sim 10{-}30$~km~s$^{-1}$ --][]{Loup93}. 
    \item Primordial, non-polluted stars move through the interstellar medium that is polluted, as in the early disc accretion scenario \citep{Bastian13}. This material is accreted {to produce discs around} lower mass stars by tidal cloud capture, disc sweep-up or  Bondi-Hoyle-Lyttleton-like accretion, {but is not efficiently accreted onto the star.} This process can occur simultaneously with (i) as long as the gas has not yet cooled to collapse and form new stars.
     \item {{This capture process can proceed more quickly than star formation, particularly if the gas reservoir is heated by the first generation of stars} \citep{Conroy11, Yaghoobi22}.}
    \item Companions form from the captured gas, and undergo eccentricity excitation due to numerous hyperbolic stellar encounters that result in accretion of the companion \citep{Hamers17, Winter22c}. 
    \item The merging of the two bodies results in angular momentum injection and mixing of the pollutant in the primary. This yields variations in stellar abundances.
    \item The primary returns to the main sequence on a thermal timescale (of order $10$~Myr). If and when it becomes an RGB star, it still retains similar surface abundances due to the fact that material is already mixed through the convective zone by the previous accretion event.
\end{enumerate}

If feasible, this model immediately has several obvious advantages over the early accretion scenario. First, the polluted star no longer has to accrete a large quantity of mass via the disc while it is still fully convective, which would require extremely rapid and efficient accretion of material on timescales shorter than a typical disc life-time. Secondly, and related to the first point, this allows stars with longer main sequence life-times (such as AGB stars) to contribute to the accreted material. Finally, it predicts a dearth of polluted stars at young ages, in line with observations. We further explore the merits and shortcomings of this model as follows.

\subsection{Origin of the contaminants}

\label{sec:origins}
\begin{figure}
    \centering
    \includegraphics[width=\columnwidth]{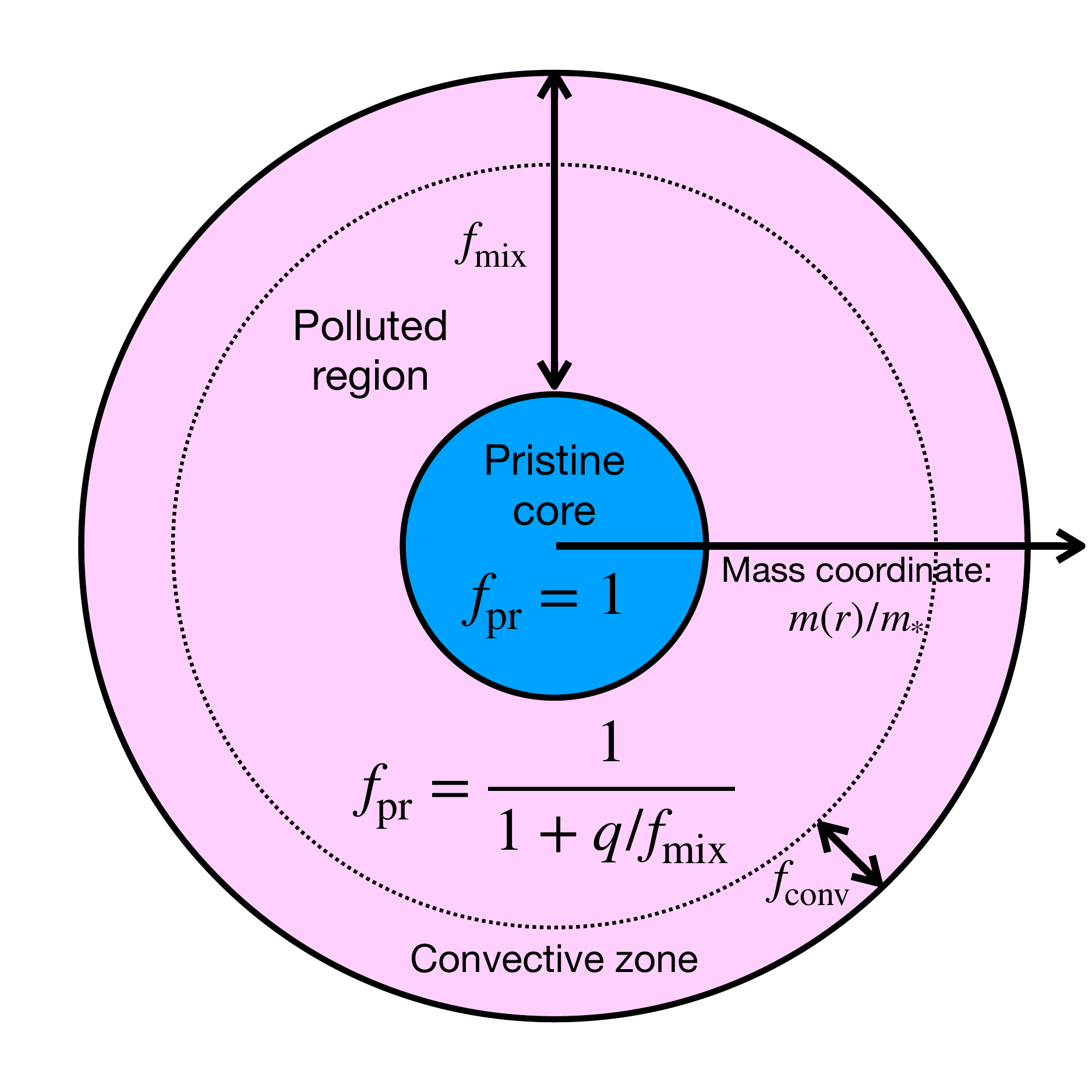}
    \caption{{Stellar pollution schematic showing our simplified model for the pollution of the star as a result of a merger with a companion of mass ratio $q$. The radial coordinate is the enclosed mass $m(r)$ as a function of radius $r$ in the star, normalised by the total stellar mass $m_*$ of the remnant. In order to parameterise the concentration of the pollutant, we assume the mixing is uniform over a fraction $f_\mathrm{mix}$ of the outer layers of the star. Generally, $f_\mathrm{mix}\geq f_\mathrm{conv}$, where $f_\mathrm{conv}$ is the mass fraction of the convective zone.} }
    \label{fig:shell_schematic}
\end{figure}

The origin of the pollutant is important for quantifying the time and amount of polluted material that can be delivered to the interstellar medium. For the model we suggest, our primary interest is in producing the requisite contamination with a sufficiently low mass ratio $q$ such that the contaminated companion is not easily detectable for the majority of stars (i.e. $q\ll 1$). Given that brown dwarf-mass companions to RGB stars appear to be common \citep{Soszynski21, Percy21}, a mass ratio of $q\sim 0.1$ is a reasonable expectation. {{It is possible to achieve the required chemical signature using arbitrarily low mass ratios if the contamination is limited to the surface layers of a star.} However, abundances of N change little along the MS and MSTO for polluted stars \citep{Briley04}, and are comparable to the polluted RGB abundances \citep{Briley97}, suggesting that much of the convective zone of stars at the tip of the RGB in 47 Tuc ($\sim 70$~percent of the stellar mass) is initially contaminated (although see discussion in Section~\ref{sec:outlook_abundances}). For simplicity, we will assume a uniform mixing of the companion material through a fraction $f_\mathrm{mix}$ of the outer stellar remnant of the merger. When observed, this results in an apparent pristine mass fraction $f_\mathrm{pr}$ at the stellar surface. {The ratio of polluted to unpolluted mass in the well mixed portion of the star is thus $q/f_\mathrm{mix}$, from which we can write:}}
\begin{equation}
    q = \frac{f_\mathrm{mix}(1-f_\mathrm{pr})}{f_\mathrm{pr}}.
\end{equation}{Figure~\ref{fig:shell_schematic} shows this parameterisation schematically.}

In the following, we ask whether possible enrichment sources can yield observed elemental abundance variations when mixed at this ratio. There are numerous potential sources for enrichment (see review by \citealt{Bastian18}), however we consider only a few relevant mechanisms here.

\subsubsection{Rapidly rotating massive stars and binaries}
\label{sec:massive}
Rapidly rotating massive stars and binaries are an attractive way to pollute a star early during its evolution. Massive main sequence (MS) stars produce many of the observed chemical abundance correlations \citep{Maeder06}. The problem in the first instance comes in allowing the fusion products to reach the surface. \citet{Decressin07} put forward mixing induced by rapid rotation as a solution. The slow wind may then be ejected into the primordial gas (allowing the required dilution) to produce a new generation of stars. However, in-of-itself it is not clear how this scenario could produce discrete populations, Mg abundance changes or overcome the mass budget problem (see Section~\ref{sec:enrfrac_mbudget}). 

A related mechanism that may produce similar chemical signatures is the interaction between massive stars in a close binary system \citep{deMink13}, which are common \citep{Sana12} and have short life-times that are comparable to that of the primordial circumstellar disc \citep[e.g.][]{Haisch01}. As an example of the abundances expected in the massive binary scenario, \citet{deMink09} computed the evolution of a $20\,M_\odot$ star in a close $15\, M_\odot$ companion with an orbital period of $12$~days. The authors show that the system sheds about $10\,M_\odot$ of material enriched in He, N, Na, and Al and depleted in C and O. The most extreme abundances for the ejecta the authors obtain are: $Y=0.64$ for He (absolute abundance) and $[X] = 1.63$, $1.44$, $0.49$, $-1.07$, $-1.13$ for Na, N, Al, C and O respectively. Here, $[X] = \log X/X_\mathrm{i}$ where $X$ is the mass fraction and $X_\mathrm{i}$ is the initial mass fraction. The Mg abundance does not change significantly, indicating that variations in its abundances \citep[and anti-correlation with Al --][]{Carretta12} must originate from another source.

{{Here we are invoking late stage accretion, rather than early disc accretion \citep{Bastian13}, to produce the companions (see Section~\ref{sec:origins})}. We are therefore no longer limited to the massive binaries that rapidly eject material on timescales $<10$~Myr, or even shorter timescales that enable the material to become convectively mixed in the pre-main sequence star \citep[see discussion by][]{Bastian18}. We are now able to revisit enrichment by other means over longer periods, such as via AGB winds that occur over several $10$~Myr timescales.}

\subsubsection{AGB stars}
\label{sec:AGB}
When less massive stars ($m_* \sim 4{-} 9 \, M_\odot$) reach the AGB phase, their ejecta can also enrich the surrounding ISM. \citet[][see also \citealt{DErcole10}]{DErcole08} envisioned a picture in which the AGB ejecta sink to core of the young globular cluster, achieve high density and undergo collapse into a second generation of stars. Variations on this mechanism are among the oldest and most popular explanations for producing the observed enrichments \citep{Cotrell81}. One reason for this popularity is that AGB stars burn H at higher temperatures than main sequence massive stars \citep[dependent on mass and metallicity][]{Prantzos07}, allowing them to activate the Al-Mg burning chain, depleting Mg and increasing Al. This goes some way to explaining the shortcomings in the binary model above. For example, \citet{Ventura09} find an $[\rm{Al}/\rm{Fe}]\gtrsim 1$, for stellar masses $m_* \gtrsim 5\, M_\odot$, bringing down the quantity of material required to produce the most extreme observed Al enhancements to $q\sim 0.3$. 

However, the composition of AGB ejecta is highly uncertain, and depends on the complex interplay between the secondary dredge-up (SDU), tertiary dredge-up (TDU) and hot bottom burning \citep[HBB; e.g.][]{Karakas14}. The ejecta must come from stars with mass $m_*\gtrsim 3.5 \, M_\odot$ to conserve the C+N+O content (we discuss in Section~\ref{sec:Bondi} that the sweep-up time of contaminants into companions occurs over $\sim 200$~Myr, such that the lower mass AGB stars do not contribute in t scenario). HBB may produce O depletion while TDU would increase the C+N+O on the surface, however the balance of these two processes alters the yields of other light elements such that it is challenging to reproduce all of the observed abundance variations. Although this might be solved by appealing to deviations in yields from theoretical predictions \citep{Renzini13}, the uncertainties in parameters required in AGB ejecta modelling make it challenging to produce quantitative predictions \citep[e.g.][]{Ventura05a, Ventura05b}. Nonetheless, AGB enrichment models require significant dilution of ejecta in pristine material, both for mass budget and abundance reasons. While the quantity of diluting material is poorly constrained, if all of the $\sim 8$~percent of the total stellar mass in AGB stars is accreted into companions around low mass stars that make up $40$~percent of the total IMF mass, then the typical mass ratio is  $\langle q \rangle \sim 0.2$. While the assumption of $100$~percent accretion efficiency is extreme, such a mass ratio would be reasonable for a binary in the context of our model. 

Apart from abundances in ejecta, a number of issues are associated with this traditional AGB enrichment scenario. In the first instance, this scenario suffers from the requirement that the gas cools sufficiently to collapse, which is only possible after $\gtrsim 100$~Myr \citep{Conroy11}. This delay is a severe problem for the AGB scenario, as the C+N+O abundance would not be conserved for AGB stars at this mass, contradicting observations \citep[e.g.][]{Dickens91}. This delay also leads to problems in retaining enough pristine gas in order for the cluster to have the necessary primordial gas for dilution, which is required for the observed chemical abundance correlations \citep[for example, a lack of variation in Li abundances -- e.g.][]{Mucciarelli11}. Related to this, without dilution the amount of mass required to produce the second generation of stars with a mass of $\sim 0.4{-}0.9$ times the mass of the initial population is far higher than the mass fraction that can be supplied by the AGB population, which is the mass budget problem that is common to many enrichment scenarios (see Section~\ref{sec:enrfrac_mbudget}). 

In the context of the companion accretion model, none of the above issues are necessarily a problem. Rather than forming a new population from collapsing gas, we require only capture of the gas (Section~\ref{sec:Bondi}). Thus the gas does not need to immediately cool to produce the next generation of stars. Dilution is naturally achieved by the mixing of the material in the polluted companion with the host star, without the need to appeal to a significant retention of primordial gas or re-accretion of fresh surrounding gas. The total quantity of mass needed is no longer determined by the total mass of the enriched stars, but the amount of mass needed to pollute a pristine star. The latter is dependent on the abundance yields from AGB stars that are highly uncertain, but given the wider time window available for pollution we can now appeal to material that originates in multiple types of progenitors (see also discussion in Section~\ref{sec:Bondi}).

\subsubsection{Internal mixing}
\label{sec:internal}
Early in the study of multiple populations in globular clusters, evolutionary mixing was suggested as the origin of the chemical inhomogeneities in RGB stars \citep{Denisenkov90}. Stellar models for RGB stars with rotational mixing are well-known to be capable of reproducing abundance anomalies for C, N, O and Na \citep[][and review by \citealt{Salaris02}]{Langer93, Charbonnel95,Denissenkov00}. However, this hypothesis was effectively dismissed with the finding that the MS and MSTO stars exhibit similar abundance patterns \citep{Cannon98, Briley04}. Since MS stars close to the turn off have small convective zones (encompassing a mass fraction $\sim 10$~percent for $m_*\approx 0.8 \, M_\odot$), enrichment could not be the product of convective mixing of inner H burning products. In addition, while the CNO cycle occurs in low mass stars, and may result in enhancements in N and depletion of C and O, variations in Na, Al, and Mg cannot by produced. Their temperatures are too low to activate the NeNa- and MgAl-chains \citep[e.g][]{Prantzos07, Prantzos17}.

However, in the scenario we put forward, a significant fraction of the evolved MS star undergoes mixing as a result of collision with the companion (Section~\ref{sec:mixing}). {Subsequently, the high angular momentum companion material induces rapid rotation inside the combined star, which may result in rotational mixing similar to those suggested to contribute to RGB surface abundances.} Thus it is possible that mixing does indeed contribute to changes in surface abundances. During companion accretion, the presence of externally produced elements {(e.g. Mg, Na and Al)} is naturally correlated with mixing of internal burning products, because a star needs to have merged with its companion to have produced the deep mixing and CNO variations. However, the heavier elements have their origin in the AGB and/or massive star ejecta. Thus, a combination of deep mixing and external origins becomes plausible \citep{Briley02}. {As commented by \citet{Briley04} on internal dredge-up versus high mass star origins for abundance variations: \textit{`we find ourselves now facing a wealth of evidence that suggests not one origin or the other but rather both'.} Indeed, the authors point to correlations between C depletion with decreasing absolute magnitude in several low-metallicity clusters \citep{Bellman01} and the C isotopic variation with stellar mass \citep{Shetrone03} as strong evidence that deep dredge-up \textit{does} contribute to abundance variations. The scenario we present in this work has the capacity to facilitate both internal mixing and massive/AGB star origins for abundance variations. }

\subsubsection{Absent supernovae ejecta}

{The small dispersion in iron content in most globular clusters indicates that the fraction of mass ejected by core collapse supernovae that is retained in the cluster cannot generally exceed a few percent \citep{Renzini13,Marino19b}. This has led some authors to conclude that the pollution must originate either before or well after the onset of these supernovae \citep{Renzini15}. {Scenarios in which the supernovae deplete the available gas reservoir while the second generation is forming may exacerbate the mass budget problem} \citep[see discussion by][]{Renzini22}. However, the mass budget problem is less severe in the companion accretion model, while in a highly structured medium supernovae ejecta may follow low density channels to escape efficiently without removing all the existing gas \citep[e.g.][]{Rogers13, Krause13}. In this work, we consider a scenario in Section~\ref{sec:compsynth} in which all massive binary ejecta are removed by core collapse supernovae, then AGB stars eject material afterwards. }

\begin{figure*}
\centering
\subfloat[\label{subfig:ONa_dilute}]{\includegraphics[width=0.5\textwidth]{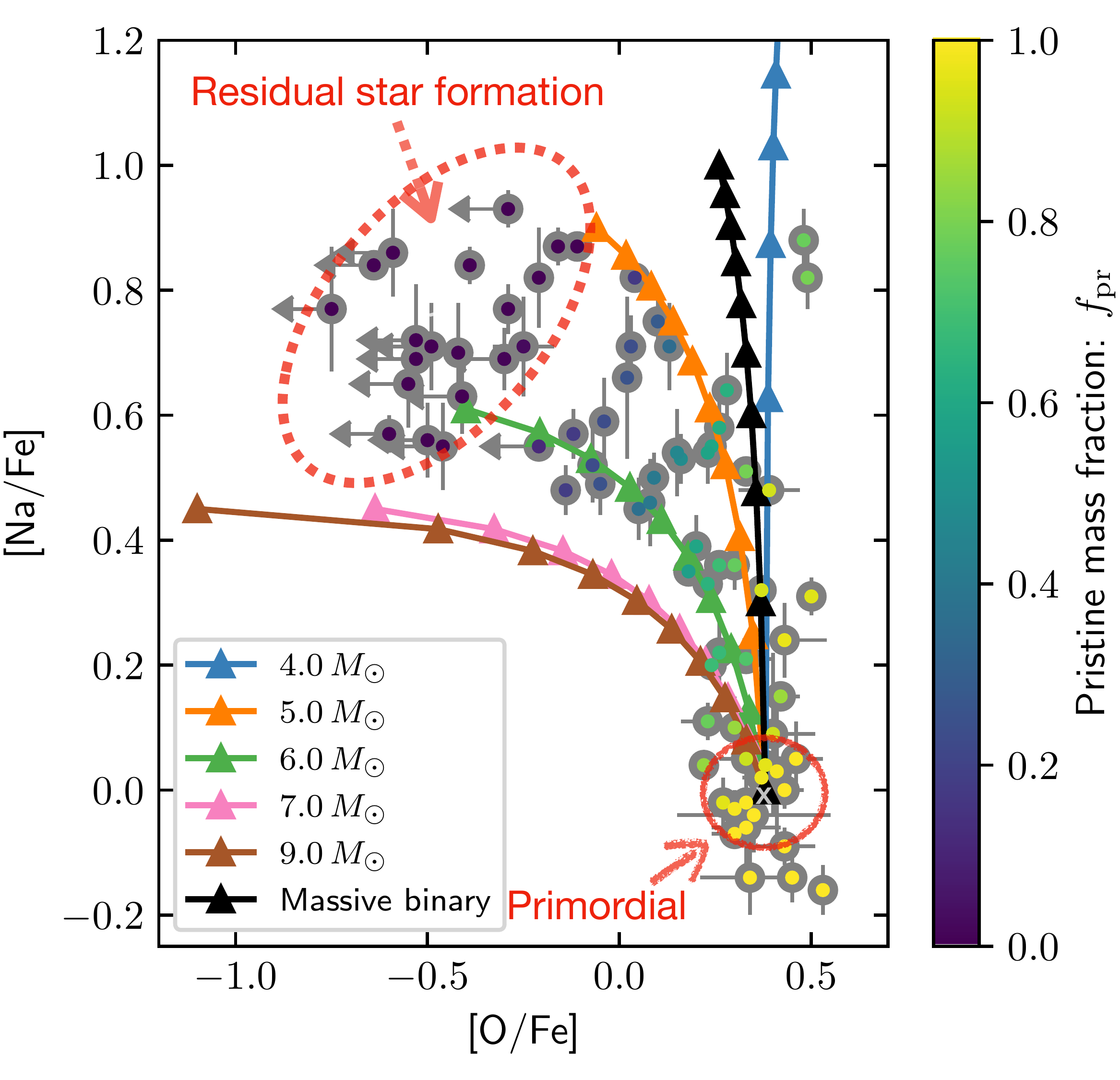}}
\subfloat[\label{subfig:fpr_q_fmix}]{\includegraphics[width=0.5\textwidth]{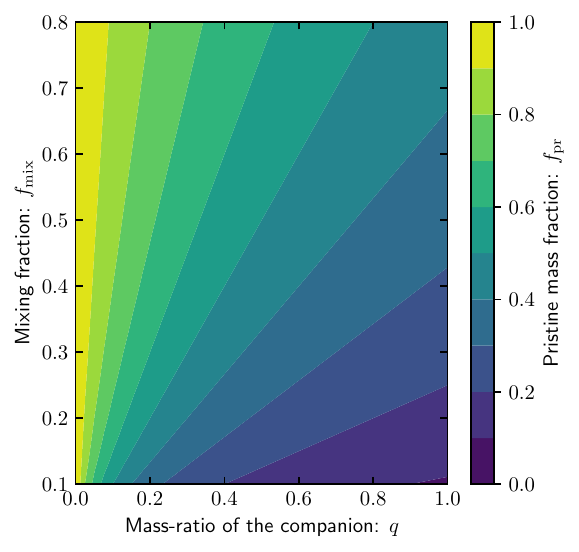}}

\caption{Dilution tracks for O and Na abundances (Figure~\ref{subfig:ONa_dilute}) and {the pristine mass fraction $f_\mathrm{pr}$ as a function of companion mass ratio $q$ and mixing fraction $f_\mathrm{mix}$ (Figure~\ref{subfig:fpr_q_fmix}).} We adopt the AGB star abundances adopted as in \citet[][coloured lines]{DErcole10} and the {average} abundances for the massive binary explored by \citet[][black lines]{deMink09}. {AGB stars of mass $m_\mathrm{AGB} \sim  4{-}9\, M_\odot$ are included because they have conserved C+N+O abundances, undergo recurrent mixing events \citep{Karakas14}, and have life-times short enough to pollute their environment before cooling and collapse of the medium \citep{Conroy11}.} The triangle markers show increments of $0.1 f_\mathrm{pr}$, where $f_\mathrm{pr}=1-f_\mathrm{poll}$ is the fraction of pristine material compared to contiminated material. We have assumed pristine abundances $\rm{[O/Fe]}= 0.33$, $\rm{[Na/Fe]}= 0.0$, $\rm{[Mg/Fe]}=0.37$ and $\rm{[Al/Fe]}= 0.11$, in order to compare with the abundances in M54 (NGC 6715) as reported by \citet{Carretta10}, shown here as circles with error bars. Upper limits are shown as arrows, points without error bars are those for which no uncertainty is listed by \citet{Carretta10}. {The colours of the observational data points show an estimate of the surface pristine mass fraction $f_\mathrm{pr}$ by minimizing the distance between this point and an interpolated AGB ejecta contour using the \textsc{scipy} package \texttt{optimize.minimize}. These values of $f_\mathrm{pr}$ can be mapped to a corresponding $f_\mathrm{mix}$ and $q$ using Figure~\ref{subfig:fpr_q_fmix}.} }
\label{fig:dilution_tracks}
\end{figure*}

\subsubsection{Outlook for producing observed abundance variations}
\label{sec:outlook_abundances}

\begin{figure}
    \centering
    \includegraphics[width=\columnwidth]{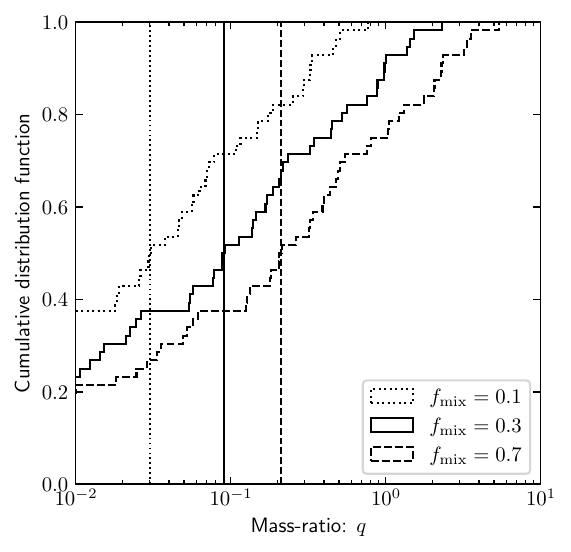}
    \caption{{The mass-ratio $q$ distribution required to produce the distribution of RGB star abundances in M54 for those stars that are not composed purely of pollutants, where purely pollutant is defined to be $f_\mathrm{pr}<0.05$. We show the distribution for fixed assumed mixing fractions $f_\mathrm{mix}=0.1$ (dotted), $0.3$ (solid) and $0.7$ (dashed), with corresponding medians shown as vertical lines. An important caveat is that the precise shape of the cumulative distribution function at the low $q$ end is dependent on the assumed primordial abundances and observational uncertainties. The shape of the function at the high $q$ end is similarly dependent on AGB model ejecta abundances and observational uncertainties. }}
    \label{fig:q_distribution_M54}
\end{figure}


We show the dilution tracks derived from the extreme abundances of the massive binary simulation of \citet{deMink09} and the AGB ejecta as adopted by \citet{DErcole10} in Figure~\ref{subfig:ONa_dilute}. We have chosen pristine abundances $[\rm{O}/\rm{Fe}] = 0.38$, $[\rm{Na}/\rm{Fe}] = 0.0$, $[\rm{Mg}/\rm{Fe}] = 0.35$ and $[\rm{Al}/\rm{Fe}] = 0.11$ in order to compare with the sample of abundances for RGB stars as reported by \citet{Carretta10}, also shown in Figure~\ref{fig:dilution_tracks}. {Dilution tracks start where the ejecta abundance lines converge, at which point the composition is $100$~percent pristine -- i.e. the fraction of pollutant fraction $f_\mathrm{poll} = 0$. As the fraction of pristine material $f_\mathrm{pr}=1-f_\mathrm{poll}$ decreases, the material becomes more concentrated until it is composed entirely of the ejecta material ($f_\mathrm{poll}=1$, $f_\mathrm{pr}=0$). We show increments of $0.1$ in $f_\mathrm{poll}$ ($f_\mathrm{pr}$) as triangles in Figure~\ref{subfig:ONa_dilute}. {We also show the pristine mass fraction $f_\mathrm{pr}$ as a function of the effective mixing fraction $f_\mathrm{mix}$ and companion mass ratio $q$ in Figure~\ref{subfig:fpr_q_fmix}.}}  

{We see in Figure~\ref{subfig:ONa_dilute} that it is challenging to produce some of the lowest O abundances. These cases imply very high concentrations of the pollutant for the most extreme variations, requiring some combination of large $q$ or small $f_\mathrm{mix}$. We suggest that this may be somewhat mitigated by a combination of a number of possible factors, including:}
\begin{enumerate}
    \item {A small number of companions with large $q$.}
    \item {Variable/inhomogeneous mixing of contaminants.}
    \item {Late collapse of the gas reservoir to form a (small) secondary population.}
    \item {Some surface enhancements of internal burning products via rotational mixing. }
    \item {Uncertainties in the model ejecta abundances. }
\end{enumerate}{For example, we indicate in Figure~\ref{subfig:ONa_dilute} how the first three points (i--iii) may shape the observed population of M54. The most extreme population can result from residual star formation, while a tail of highly polluted stars may result from a range of $f_\mathrm{mix}$ and $q$ (see Figure~\ref{subfig:fpr_q_fmix}).}

{In Figure~\ref{fig:q_distribution_M54} we show the distribution of the mass-ratios $q$ required to produce the inferred pristine mass fractions, assuming fixed mixing fractions $f_\mathrm{mix}$. Here we exclude those stars composed entirely of pollutant ($f_\mathrm{pr}<0.05$), which we assume are the results of star formation directly from the ejecta. We include all remaining stars, although we highlight that within the companion accretion model we expect some fraction ($\sim 20$~percent) of companions to have been ionised rather than accreted. We therefore expect to somewhat underestimate the median $q$ when including all of the RGB stars. With this caveat, the median $q$ varies between $0.03$ and $0.2$ for $f_\mathrm{mix}$ between $0.1$ and $0.7$. These values would be consistent with the available mass from massive and/or AGB stars, which contribute up to $25$~percent and $20$~percent of the mass of the low mass stellar population respectively. However, if $f_\mathrm{mix}=0.7$, we also apparently require a tail of high mass ratio companions up to $q\sim 4$, which may be challenging to produce via the companion accretion model. We highlight that the exact $f_\mathrm{pr}$ in the low $f_\mathrm{pr}$ limit strongly depends on the AGB or massive binary model abundances, which remain uncertain. However, if the ejecta abundances are accurate, we may still produce the low $f_\mathrm{pr}$ stars by adopting small $f_\mathrm{mix}$.}

{Some empirical constraints limit the degree of variation in the mixing fraction $f_\mathrm{mix}$. Empirically, in 47 Tuc and M71 \citet{Briley04} find similar nitrogen and carbon abundances along the MS and MSTO stars, which have relatively thin convective zones, spanning a range in mass fraction $f_\mathrm{conv}\sim 0.01-0.1$. This would be consistent with any mixing fraction $f_\mathrm{mix} \gtrsim \max (f_\mathrm{conv})\sim 0.1$. A more stringent constraint follows that the abundance variations are also approximately constant among polluted RGB stars, as surveyed by \citet{Briley97}. In this sample, $f_\mathrm{conv}$ varies up to $\sim 70$~percent, suggesting that $f_\mathrm{mix}$ is not less than an order of magnitude lower than this value across the sample. {However, some variation of a factor few in $f_\mathrm{mix}$ over this range is not ruled out \citep[see discussion in Section~4.3 of][]{Briley97}.}}

{Residual star formation from the pollutant in some clusters also remains a viable way to produce the highly polluted population, as in previous models for self-enrichment \citep[e.g.][]{DErcole10}. In this scenario, the second population would be present early in the cluster lifetime. This apparently contradicts the absence of multiple populations in young clusters \citep[e.g.][]{Martocchia19} and the low age spreads in slightly older clusters with multiple populations \citep[e.g.][]{Martocchia18b}. However, such a mechanism would be contingent on retaining a sufficient quantity of gas at the time star formation is instigated. This may explain why clusters with large mass-radius ratio $M_\mathrm{c}/R_\mathrm{c}$, such as M54 and NGC 2808 ($M_\mathrm{c}/R_\mathrm{c} = 2.4 \times 10^5 \,M_\odot \,\mathrm{pc}^{-1}$ and $3.4 \times 10^5 \,M_\odot \,\mathrm{pc}^{-1}$ respectively) {exhibit extreme abundance variations not seen in} 47 Tuc ($M_\mathrm{c}/R_\mathrm{c} =1.3 \times 10^5 \,M_\odot \,\mathrm{pc}^{-1}$, see discussion in Section~\ref{sec:compsynth}). The young clusters surveyed for multiple populations are not so massive or dense as the most massive (old) globular clusters, particularly when factoring in dynamical mass-loss. If clusters less concentrated than 47 Tuc {cannot form a population composed purely of AGB ejecta} then this scenario appears consistent with observations \citep{Yaghoobi22}. However, this still requires that in some clusters that $\sim 20$~percent of the present day cluster mass forms from pure ejecta, and the same in companions. If massive binary and AGB ejecta can both contribute to this budget, then a maximum of $45$~percent of the present day low mass population can be reached. This is sufficient, and can be effectively enhanced by the concentration of the ejecta material into the core of the cluster \citep[e.g.][]{Calura19}, where stars survive the subsequent dynamical ejection stars over Gyr timescales.  }

{There remain some further concerns in appealing to self-enrichment by massive progenitors. In particular, some authors predict that AGB stars do not give Na variations, which would rule them out as candidates for these variations \citep[e.g.][]{Doherty14}. Predicted yields from all enrichment sources may struggle to reproduce observed variations when more than two elements are included \citep{Bastian15b}. For example, He abundance variations are often over produced when sufficient material is injected to produce the observed O depletion and Na enhancement. We will discuss this issue more quantitatively the role of the pollutant in producing observed abundance variations in Section~\ref{sec:compsynth}. }


\subsection{Companion formation from contaminants}
\label{sec:Bondi}

\subsubsection{General principle}
\label{sec:gas_reservoir}

{We now consider how companions with mass ratio $q\sim 0.1$ may form around stars in the globular cluster. Unlike the early disc scenario \citep{Bastian13}, we can now consider a slower accumulation of material over $100$~Myr timescales, rather than the $\ll 10$~Myr needed to pollute a star while it is still convective. The collision of the companion with the host will supply the mixing (see discussion in Section~\ref{sec:mixing}). We first consider the properties of the gas reservoir, and then some different mechanisms that may allow a star to gain a companion composed of massive binary or AGB ejecta.} 

{In all scenarios, we consider an ISM that is the product of slow stellar winds that are unable to escape the potential of the dense cluster. These winds may occupy the core of the cluster \citep{DErcole08}, but do not undergo rapid collapse to produce a second generation \citep{Conroy11}. The latter authors argue that material does not collapse while the Lyman-Werner photon flux remains sufficient to prevent the formation of molecular hydrogen at typical interstellar distances within dense clusters (i.e. a few $100$~Myr), maintaining a temperature of $\gtrsim 100$~K. However, we will show in Section~\ref{sec:acc_outlook} that, even if star formation proceeds similarly to normal molecular clouds, companions can feasibly be produced on shorter timescales.} The density for the contaminant reservoir can be simply estimated as:
\begin{equation}
   \rho_\mathrm{gas} =  f_\mathrm{wind} \cdot \frac{ 3 M_\mathrm{c,0}}{4\pi R_\mathrm{gas}^3} = \frac{ 3 M_\mathrm{gas}}{4\pi R_\mathrm{gas}^3}
\end{equation}where $f_\mathrm{wind} \sim 0.1$ is the fraction of material re-injected into the ISM by the AGB and massive stars, $M_\mathrm{c,0} \sim 10^6 \, M_\odot$ is the initial stellar mass of the cluster, and $R_\mathrm{gas} \sim 1$~pc is the radius in which the gas is contained (of order the core radius). These numbers yield a density $\rho_\mathrm{gas} \sim  3\cdot 10^{-18}$~g~cm$^{-3}$. 

In the framework we present in this work there is no inherent constraint on whether the pollutant is from AGB or massive (binary) stars. If the first wave of ejecta from massive stars is very dense \citep[e.g.][]{Wunsch17} it may result in rapid accretion onto discs/companions around young stars or it may survive the feedback due to preferential clearing of the gas along low density regions between filaments \citep[e.g.][]{Dale14}. We now consider different mechanisms by which the ejecta can be captured onto the low mass stellar population. 


\subsubsection{Formation of companion in a disc}
{We generally consider a scenario in which the first generation of stars capture the gas to produce a circumstellar disc. {If the captured gas cannot cool immediately, it will form a structure that is initially supported by pressure and rotation.} This disc is gravitationally unstable if $Q\lesssim 1$, where $Q$ is the \citet{Toomre64} parameter:}
\begin{equation}
\label{eq:toomreQ}
    Q = \frac{\kappa c_\mathrm{s}}{\pi G \Sigma}
\approx  q^{-1} \frac{c_\mathrm{s}}{v_\mathrm{K}},
\end{equation}
{where $\kappa\approx \Omega_\mathrm{K} = R_\mathrm{disc} v_\mathrm{K}$ is the epicyclic frequency, similar to the Keplerian frequency $\Omega_\mathrm{K}$, $\Sigma \approx M_\mathrm{disc} /\pi R_\mathrm{disc}^2$ is the mass of the accreted contaminant, and $c_\mathrm{s}$ is the sound speed to the accreted gas when it fragments. We have adopted $q = M_\mathrm{disc}/m_*$ in the RHS of equation~\ref{eq:toomreQ}. If the gas is able to cool to $c_\mathrm{s} = 1$~km~s$^{-1}$ (temperature $T\sim 200$~K), then for $v_\mathrm{K} \sim 10$~km~s$^{-1}$ and $q\gtrsim q_\mathrm{crit} =  0.1$ the gas is gravitationally unstable and collapses to form a (sub-)stellar companion, or companions.}

{{If a large quantity of gas is captured at once (for example, via tidal capture), then the gravitational collapse may initiate immediately without the need for long-lived disc survival. However, a replenished gaseous disc composed of ejecta material may also survive longer than typical protoplanetary discs} \citep[$ 3{-}10$ Myr -- e.g.][]{Haisch01}. For a low mass main sequence star, the X-ray and EUV luminosity of the host star drops rapidly over time \citep{Tu15}, such that the internal photoevaporation of the disc becomes less efficient \citep[e.g.][]{Alexander06, Owen10}. External photoevaporation of protoplanetary discs by neighbouring stars is dominated by massive stars of mass $\gtrsim 30\,M_\odot$, absent after several Myr  \citep[e.g.][see \citealt{Winter22d} for a review]{Johnstone98}. Accretion may be suppressed for the disc around a main sequence star when the spherical Alfven radius exceeds the co-rotation radius \citep{Konigl91, Armitage95, Clarke95}. This suppression may be efficient for stars that rotate rapidly, particularly if they lose their primordial discs early due to external photoevaporation \citep[e.g.][]{Clarke00, Roquette21}. Thus the disc lifetime need not be prohibitive for mass accumulation.}

\subsubsection{Bondi-Hoyle-Lyttleton accretion}

Material may accrete slowly via BHL style accretion in a scenario similar to that considered by \citet{Throop08} and \citet{Moeckel09}. {{When the gas medium is not homogeneous, we expect the stagnation point to be offset from the axis that is coincident with the velocity vector of the star.} Material accreted in this way retains angular momentum, and so forms a disc that is not accreted immediately onto the star.} We can estimate the average rate at which a star captures material in such a scenario using the \citet{Bondi52} accretion rate \citep[see also][]{Shima85}:
\begin{equation}
    \dot{M}_\mathrm{BHL} \approx \int_0^\infty\pi R_\mathrm{BHL}^2 \rho_\mathrm{gas} (\lambda c_\mathrm{s}^2 +v_*^2)^{1/2} \, g(v_*) \,\mathrm{d}v_*, 
\end{equation}where $\lambda = e^{3/2}/4 \approx 1.12$ in the isothermal limit,  $g(v_*)$ is the probability that a star is moving with velocity $v_*$ through the gas and
\begin{equation}
    R_\mathrm{BHL} = \frac{2 G m_*}{c_\mathrm{s}^2+v_*^2}
\end{equation}is the BHL radius for a star of mass $m_*$, and the gas in the interstellar medium has sound speed $c_\mathrm{s}$. Since $g \approx v_*^2 /2\sqrt{\pi}\sigma_v^3$ where $v_*<\sigma_v$:
\begin{equation}
\label{eq:Mdot_cont}
    \dot{M}_\mathrm{BHL} \approx \frac{2\sqrt{\pi}G^2 m_*^2 \rho_\mathrm{gas}}{\sigma_v^3} \left[\mathcal{I}(\sigma_v)-\mathcal{I}(0)\right]
\end{equation}where the term in square brackets on the RHS is defined by the function:
\begin{multline}
    \mathcal{I}(x) = -\frac{x \sqrt{c_\mathrm{s}^2\lambda + x^2}}{2 (c_\mathrm{s}^2 + x^2)} +\\ \ln\left(\sqrt{c_\mathrm{s}^2 \lambda + x^2} + x\right) + \frac{(\lambda - 2) \tan^{-1} \left(\frac{x\sqrt{\lambda - 1} }{\sqrt{c_\mathrm{s}^2 \lambda + x^2}}\right)}{2 \sqrt{\lambda - 1}}. 
\end{multline}We can adopt the sound speed $c_\mathrm{s} \sim 1$~km~s$^{-1}$, $v_* \sim 10$~km~s$^{-1}$ and $m_* = 1 \, M_\odot$, we find $R_\mathrm{BHL} \approx 18$~au. With a similar $\sigma_v=10$~km~s$^{-1}$, then we obtain an average $\langle \dot{M}_\mathrm{BHL}\rangle \sim  6\cdot  10^{-9} \,M_\odot$~yr$^{-1}$. This results in an average change in the mass ratio of the contaminant to the pristine material in the whole system $\dot{q} \sim 6\cdot 10^{-3}$~Myr$^{-1}$. A mass ratio $q \sim 0.1$ is therefore reached (globally averaged) on a timescale of tens of Myr, shorter than the several $100$~Myr that may be required for the Lyman-Werner flux density to drop to a level to allow the ejecta to cool and rapidly form a second population of stars \citep{Conroy11}. {However, we revisit this rate in Section~\ref{sec:acc_outlook}, showing that BHL accretion is generally less efficient than other gas capture mechanisms.}

\subsubsection{Tidal cloud capture}

{In a highly sub-structured medium populated by dense
cloudlets, gas can be accreted onto pre-existing stars via the mechanism of tidal capture and disruption of passing cloudlets \citep[see][]{Dullemond19, Kuffmeier20}. This might work similarly to the tidal compression (and disruption) between the black hole and an infalling cloud in the galactic centre, subject both to the pressure from the low density ambient medium and the tidal forces of the compact object as modelled by a number of authors \citep{Burkert12,Lucas13, Steinberg18}. The disruption of the cloud and capture of gas in this way may be responsible for bursts of star formation in the galactic centre \citep{Bonnell08}, possibly enhanced by convergent flows \citep{Hobbs09}.}

{To attempt a quantitative estimate at the rate of tidal capture, here we speculate that this capture may operate similarly to the tidal exchange of orbital energy during close passages between stars and other stars \citep[e.g.][]{Rob68,Fab75, Bon03, Winter22a} or protoplanetary discs \citep{Ostriker94, Winter18}. If we assume that the cloud capture is physically similar to stellar capture, then the capture rate in the hyperbolic regime is \citep{Winter22a}}:
\begin{multline}
\label{eq:Gamma_capt}
    \Gamma_\mathrm{capt} \sim  0.013\left( \frac{n_\mathrm{clouds}}{10^5 \,\mathrm{pc}^{-3}}\right) \, \left[ \left( \frac{R_\mathrm{cloud}}{10\,\mathrm{au}}\right)^5\left( \frac{M_\mathrm{cloud}}{1\,M_\odot}\right) \right.
    \\ \left. \times \left( \frac{\sigma_v}{10\,\mathrm{km s}^{-1}} \right) \frac{m_* (M_\mathrm{cloud}+m_*)}{M_\mathrm{cloud}^2 } \right]^{1/3} \,\mathrm{Myr}^{-1}.
\end{multline}
{Here $n_\mathrm{clouds} = \rho_\mathrm{gas}/M_\mathrm{cloud}$ is the number density of clouds. In the hyperbolic (high velocity) regime, the capture cross-section is insensitive to the internal equation of state \citep{Lee86}, such that equation~\ref{eq:Gamma_capt} might be a reasonable approximation for tidal capture by clouds. }

{The assumption of an idealised geometry as well as uncertainties in cloud properties mean that equation~\ref{eq:Gamma_capt} only offers an order of magnitude estimate. In addition, tidal encounters involving `capture' of a gas cloud probably result in tidal disruption of the cloud rather than capture of the entire mass. Therefore, to estimate a mass accretion rate using equation~\ref{eq:Gamma_capt}, we multiply the amount of mass expected to be gained in each capture encounter:}
\begin{equation}
    \dot{M}_\mathrm{capt} =  f_\mathrm{capt} \Gamma_\mathrm{capt}  M_\mathrm{cloud} .
\end{equation}{All of the complex physics is hidden in the factor $f_\mathrm{capt}$. For illustrative purposes, we will here assume that the capture process can yield a maximum efficiency of $f_\mathrm{capt}= \min \{0.1 m_*/M_\mathrm{cloud}, 1\}$, such that a maximum of $q= 0.1$ is produced in one encounter.}

\subsubsection{Disc sweeping}

{Early disc sweeping is the scenario for surface pollution envisioned by \citet{Bastian13}. This mechanism may still operate in the context of the companion accretion model. If the primordial disc (or a subsequently formed one) is continuously supplied with gas, then this can enhance the encounter cross-section and lead to rapid sweep-up of material. Integrating over the relative velocities, the accretion rate for disc of radius $R_\mathrm{disc}$ and mass ratio $q$ is \citep{Binney08}:}
\begin{multline}
\label{eq:Mdot_sweep}
    \dot{M}_{\mathrm{sweep}} =  16 \sqrt{\pi} \rho_\mathrm{gas} \sigma_v R_\mathrm{disc}^2 \left(1 + \frac{G m_*(1+q)}{2\sigma_v R_\mathrm{disc}} \right) \\
    = 1.63 \times 10^{-8} \chi_\mathrm{grav} \left(\frac{M_\mathrm{gas}}{10^5 \, M_\odot} \right) \\ \times \left(\frac{R_\mathrm{gas}}{1\,\mathrm{pc}}  \right)^{-3} \left(\frac{\sigma_v}{10 \,\mathrm{km}\,\mathrm{s}^{-1}} \right)\left( \frac{R_\mathrm{disc}}{10\, \mathrm{au}}\right)^2  \, M_\odot \, \mathrm{yr}^{-1}.
\end{multline}{Here we define the gravitational focusing parameter:}
\begin{equation}
    \chi_\mathrm{grav} = 1+ 0.44 \left( \frac{m_*(1+q)}{1\,M_\odot}\right) \left( \frac{\sigma_v}{10 \,\mathrm{km s}^{-1}}\right)^{-1} \left(\frac{R_\mathrm{disc}}{10\,\mathrm{au}} \right)^{-1}
\end{equation}
{We have adopted the average density of the gas reservoir $\rho_\mathrm{gas} = 3M_\mathrm{gas}/4\pi R_\mathrm{gas}^3$, where $M_\mathrm{gas}$ is the total mass and $R_\mathrm{gas}$ is the radial extent. We see that the disc does not need to be very extended in order to yield rapid accretion if the density is large. In dense environments the initial disc extent may shrink due to star-disc encounters and external heating \citep[e.g.][]{Winter18b}. Perhaps more importantly in a dense interstellar medium, such a disc is also subject to ram pressure stripping. Here we adopt $R_\mathrm{disc}= 10$~au, for which a disc survives face-on accretion in a medium with density $\sim 10^{-18}$~g~cm$^{-3}$ moving at a mutual velocity of $\sim 10$~km~s$^{-1}$ \citep{Wijnen17}.  }

\subsubsection{Accretion after companion formation}

{Once a companion is formed, either from primordial or ejecta material, some further accretion may occur due to angular momentum exchanges between the gas and the companion. Whether residual infalling material is more likely to accrete on to the host star or the companion depends on the specific angular momentum and temperature of the gas \citep[e.g][]{Artymowicz83, Bate97a, Bate97b}, though this issue has not been investigated in the context of BHL accretion. If a circumbinary disc can form in this way, for low $q$ we might expect the majority of this disc mass to accrete onto the secondary \citep{Duffell20}, unless the disc collapses to form another companion.} 

\subsubsection{Overall rates and outlook for companion formation}
\label{sec:acc_outlook}

\begin{figure}
    \centering
    \includegraphics[width=\columnwidth]{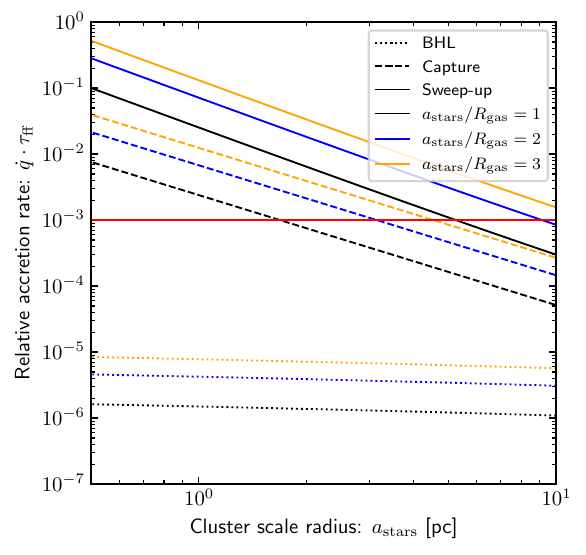}
    \caption{{The relative accretion rate (change of mass ratio multiplied by the free-fall time) according to BHL style accretion (dotted lines), tidal cloud capture (dashed lines) and disc sweep-up (solid lines) for a solar mass star as a function of the scale radius of the stellar population $a_\mathrm{stars}$. We adopt a total initial mass $M_\mathrm{c,0}=10^7\, M_\odot$, of which $40$~percent remains in the low mass stellar population and $10$~percent goes into the gas reservoir ($M_\mathrm{gas}=10^6 \, M_\odot$). The gas radius $R_\mathrm{gas}$ that is equal to (blue), one half (blue) and one third (yellow) of $a_\mathrm{stars}$. The red horizontal line shows the threshold above which we expect companion formation to yield $q\sim 0.1$ before significant star formation has occurred, assuming the star formation efficiency per free-fall time is $\epsilon_\mathrm{SF}= 0.01$.}}
    \label{fig:qdot_Rgas}
\end{figure}

{We now estimate the timescale of gas capture relative to the free-fall timescale of the gaseous reservoir:
\begin{equation}
    \tau_\mathrm{ff} = \sqrt{\frac{3\pi}{32 G \rho_\mathrm{gas}}},
\end{equation}for gas density $\rho_\mathrm{gas}$. Here, for simplicity, we assume a Plummer density profile for the stars:}
\begin{equation}
    \rho_\mathrm{stars}(r) = \frac{3 M_\mathrm{stars}}{4\pi a_\mathrm{stars}^3} \left(1+\frac{r^2}{a_\mathrm{stars}^2} \right)^{-5/2}
\end{equation}{as a function of radius $r$ within the cluster. The three dimensional velocity dispersion is then:}
\begin{equation}
    \sigma_v(r) = \sqrt{\phi/2},
\end{equation}where 
\begin{equation}
    \phi(r) = \frac{G M_\mathrm{stars}}{a_\mathrm{stars}} \left(1+\frac{r^2}{a^2}\right)^{-1/2}.
\end{equation}{We assume a uniform gas density profile inside a sphere of radius $R_\mathrm{gas}$, which we consider to be proportional to $a_\mathrm{stars}$. We consider scenarios in which the ejecta are more concentrated than the stellar population, which might plausibly arise when the high mass progenitors are mass segregated \citep[as discussed by][]{Bastian13, Calura19}. Where appropriate, we assume that the ejecta is composed of clumps of gas with radius $R_\mathrm{cloud}=10$~au and mass $M_\mathrm{cloud}=0.1\,M_\odot$, which is marginally Jeans stable at temperature $T=200$~K. } 

{In Figure~\ref{fig:qdot_Rgas} we show  $\dot{q}\cdot \tau_\mathrm{ff}$, the product of the free-fall timescale and the rate of change of the mass ratio $q$ for a disc around a solar mass star undergoing BHL, tidal cloud capture or disc sweep-up at the centre of a cluster with scale radius $a_\mathrm{stars}$. If we require $q$ to reach $0.1$ over the timescale for star formation:}
\begin{equation}
    \tau_\mathrm{SF} = \epsilon_\mathrm{SF}^{-1} \tau_\mathrm{ff}, 
\end{equation}{{where} $\epsilon_\mathrm{SF}$ is the star formation efficiency per free-fall time. The value of $\epsilon_\mathrm{SF}$ is generally estimated to be $\lesssim 1$~percent \citep[see discussion by][]{Krumholz19}, in which case configurations above the red horizontal line in Figure~\ref{fig:qdot_Rgas} would produce sufficient companion mass before significant star formation. If $\epsilon_\mathrm{SF}$ is smaller due to the increased temperature of the gas \citep{Conroy11}, then this red line would move downwards. In general, tidal capture and sweep-up are promising mechanisms by which stars can capture the requisite pollutant to produce a companion, while BHL accretion may be too slow.}

\subsection{Survival of the gas}

{For a significant fraction of polluted material to end up in companions that are subsequently accreted, we require that the gas reservoir in the core of the cluster is maintained until the gas can be captured. We now investigate whether this is consistent with the absence of evidence of cold dust and gas in young massive clusters \citep{Bastian14, CabreraZiri15}, nor significant quantities of fully ionised hydrogen indicating ongoing star formation \citep{Bastian13b}. This absence has been attributed to the removal of gas via ram pressure stripping as a young globular cluster passes through a dense medium \citep{Chantereau20}. As discussed by \citet{Longmore15}, the absence of observational signatures of residual gas in young massive clusters represents a significant problem for traditional models of second generation formation.}

\subsubsection{Gas mass evolution}
\label{sec:rapid_acc}

{We can estimate the evolution of the mass of the gas reservoir in the cluster over time by comparing the rate of accretion onto the circumstellar discs (we assume disc sweep-up here, using equation~\ref{eq:Mdot_sweep}) and the approximate rate at which material is ejected into the medium:}
\begin{equation}
    \dot{M}_\mathrm{eject} = \begin{cases} \frac{M_\mathrm{c,0}}{\langle m_* \rangle} |\dot{m}_\mathrm{TO}| \xi(m_\mathrm{TO}) \Delta  m_{\mathrm{TO}} \qquad & m_\mathrm{TO} >m_\mathrm{min}\\
    0 \qquad & m_\mathrm{TO} < m_\mathrm{min}
    \end{cases}.
\end{equation}{Here $\Delta m_{\mathrm{TO}}= m_\mathrm{TO}-m_\mathrm{rem}$ is the mass of the stars at main sequence turn-off $m_\mathrm{TO}$ minus the remnant mass $m_\mathrm{rem}$. We will assume $m_\mathrm{rem}\ll m_\mathrm{TO}$ to yield maximal mass ejection. We have introduced $\xi$ which is the mass function, for which in the high mass limit we will assume $\xi \propto m^{-2.35}$, normalised to yield $8$~percent of the total stellar mass in the AGB stars in the mass range $4{-}9\, M_\odot$, where the minimum mass of a star contributing to the gas reservoir is $m_\mathrm{min}=4\,M_\odot$. The normalisation is given by the number of stars in the cluster or the total initial stellar mass $M_\mathrm{c,0}$ divided by the average stellar mass $\langle m_* \rangle \approx 0.5\,M_\odot$. We assume that the main sequence life-time for a star of mass $m_*$ is}
\begin{equation}
    \tau_\mathrm{MS} = 20 \left( \frac{m_*}{9\, M_\odot} \right)^{-2.5} \, \mathrm{Myr},
\end{equation}such that
\begin{equation}
   \dot{m}_\mathrm{TO} =  -0.18 \left(\frac{t}{20\,\mathrm{Myr}}\right)^{-7/5} \, M_\odot \, \mathrm{Myr}^{-1}.
\end{equation}{We adopt a maximum mass $m_\mathrm{max}=50\, M_\odot$ that contributes to the ejecta, although our results are not sensitive to this choice.}

{Using the above, the rate of change of the total mass of gas in the reservoir is then:}
\begin{equation}
\label{eq:Mdot_gas}
    \dot{M}_\mathrm{gas} =  \dot{M}_\mathrm{eject} - f_\mathrm{lm} M_\mathrm{c,0} \langle \dot{q}_\mathrm{sweep}\rangle - \dot{M}_\mathrm{SF}.
\end{equation}{Here we estimate $f_\mathrm{lm}\approx 0.4$, the fraction of the initial stellar mass in the low mass stars ($m_* \lesssim 1 \, M_\odot$) that capture the gas \citep[with an IMF following][]{Kroupa01}. We will here assume that the total star formation rate $\dot{M}_\mathrm{SF}=0$. The rate of change of the average disc mass-ratio is:}
\begin{equation}
\label{eq:qdot}
    \langle \dot{q}\rangle =  \langle \dot{q}_\mathrm{sweep}\rangle 
 = \dot{M}_\mathrm{sweep}/\langle m_* \rangle.
\end{equation}

{The only further information needed is the total initial stellar mass of the cluster and the scale radius of the stars, $a_\mathrm{stars}$, and radius of the gas reservoir $R_\mathrm{gas}$. Hereafter, we assume a Plummer sphere density profile of stars, and a uniform gas density such that $R_\mathrm{gas}=a_\mathrm{stars}$. We will assume that the half-mass of the present day cluster is equal to the half-mass of the initial cluster $R_\mathrm{hm}$, such that $a_\mathrm{stars} \approx R_\mathrm{hm}/1.3$. For the total stellar mass, unless otherwise stated we will assume the present day low mass star population has been dynamical depleted by 50~percent, while the total mass in low mass stars is $40$~percent. Hence the initial total mass is $M_\mathrm{c,0}  = 5 M_\mathrm{c}$, where $M_\mathrm{c}$ is the present day cluster mass. }

\begin{figure}
    \centering
    \includegraphics[width=\columnwidth]{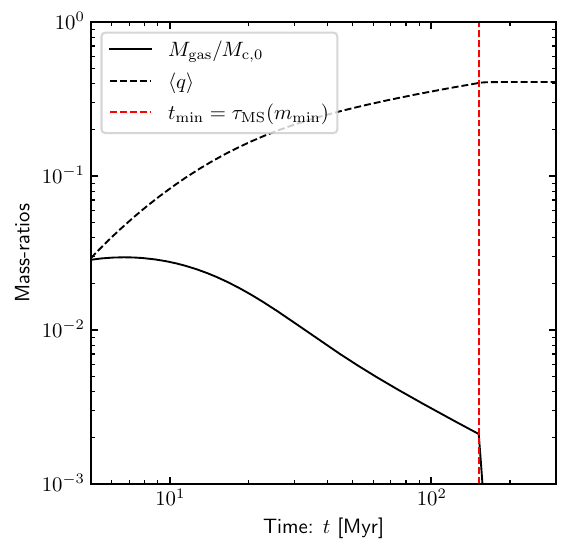} \caption{{ Evolution of the fraction of the cluster mass remaining in ejected material that has not been swept up by discs $M_\mathrm{gas}/M_\mathrm{c,0}$,
     together with the averaged disc mass ratio $\langle{q}\rangle$. We adopt a simple model for NGC 2808: $M_\mathrm{c,0} = 4.3 \times 10^6 \, M_\odot$, $a_\mathrm{stars}=R_\mathrm{gas}=3.7$~pc (see text for details). We show the fraction of the total mass ejected from massive binaries (left of red line) and AGB stars (right of red line) as a solid black line. We assume accretion onto a disc with $R_\mathrm{disc}=10$~au around a star with $m_* = 0.5\,M_\odot$ according to equation~\ref{eq:Mdot_sweep}. }  }
    \label{fig:ejecta_mass}
\end{figure}

{In order to demonstrate the evolution of the gas content and disc mass ratio, we consider parameters appropriate for NGC 2808. We adopt $M_\mathrm{c}=8.6 \times 10^5$ and $R_\mathrm{hm}=3.9$~pc \citep{Hilker20}. We integrate the coupled ordinary differential equations~\ref{eq:Mdot_gas} and~\ref{eq:qdot} using a fourth order Runge-Kutta scheme. The results are shown in Figure~\ref{fig:ejecta_mass}. We find that in this case, the mass in the gas remains $<1$~percent of the total stellar mass at time $t\gtrsim 30$~Myr. This is similar to the constraint quoted by \citet{Bastian14} for several clusters aged $\sim 30{-}300$~Myr using Spitzer $160$~$\mu$m observations. While the upper limit for the dust mass is more stringent if the dust is hot, we also expect a high optical depth at these wavelengths, which we revisit below. }

\subsubsection{Extinction and dust}
\label{sec:slow_acc}
{Even if the accretion of the gas reservoir is slower than the above estimate, the properties of the medium may further hide observational signatures of a massive gas reservoir. With regards to the cold medium, the bright sub-mm emission and internal extinction is geometry dependent. \citet{Bastian14} and \citet{Longmore15} argue that the lack of FIR dust emission and the lack of (optical) colour gradients in  the stellar population point to an absence of dust in the interstellar medium. However, this absence may be explained by the turbulent properties of the medium. If this medium is composed of cloudlets of radius $R_\mathrm{cloud}$ and masses $M_\mathrm{cloud}$, then the optical depth of the dust in a cloudlet is:}
\begin{align}
    \tau &\approx \kappa_\nu \Sigma_\mathrm{dust} \\ 
    &\approx 90 \left(\frac{\kappa_\nu}{10 \, \rm{cm}^2 \,\rm{g}^{-1}} \right) \left(\frac{M_\mathrm{cloud}}{0.1\, M_\odot} \right) \left( \frac{\rho_\mathrm{dust}}{10^{-2} \cdot \rho_\mathrm{gas}}\right)\left( \frac{R_\mathrm{cloud}} {10~\rm{au}}\right)^{-2},
\end{align}{where the dust-to-gas ratio is $\rho_\mathrm{dust}/\rho_\mathrm{gas}$ and we estimate the opacity:}
\begin{equation}
    \kappa_\nu = \kappa_0 \left(\frac{\nu}{\nu_0} \right)^{\beta},
\end{equation}{where $\beta \approx 1$ and $\kappa_\nu \sim 10$~cm$^{2}$~g$^{-1}$ at $10^3$~GHz \citep{Hildebrand83, Beckwith90}. Clearly, at $160$~$\mu$m, the longest wavelength observed with \textit{Spitzer} by \citet{Bastian14}, this yields a large optical depth. For cloudlets with radius $R_\mathrm{cloud} \sim R_\mathrm{BHL} \sim 10$~au and mass $M_\mathrm{cloud} \sim 0.1 \, M_\odot$, this results in an underestimate by a factor $>100$ in the observed total gas mass at $160$~$\mu$m. The total flux of the warm dust is therefore dependent on the amount of mass in small cloudlets. } 

{In the case of a clumpy medium, \citet{Longmore15} also predicts a patchy extinction pattern. However, this is contingent on the effective (two dimensional) filling factor of the gas clouds. This can be approximated:}
\begin{equation}
    \mathcal{F} \approx \frac{M_\mathrm{gas}} {M_\mathrm{cloud}}\left( \frac{R_\mathrm{cloud}}{R_\mathrm{gas}} \right)^{2}
\end{equation}{if $\mathcal{F}\ll 1$. Factoring in three dimensional structure, the average fraction of obscured stars in the cluster is always $<\mathcal{F}$. If we assume $M_\mathrm{gas} = 10^5\, M_\odot$ and $M_\mathrm{cloud} =0.1 \, M_\odot$, with $R_\mathrm{cloud} = 10$~au and $R_\mathrm{gas} = 1$~pc, we obtain a small $\mathcal{F} \approx 2 \times 10^{-3}$. This is a conservative estimate in terms of the reservoir properties; the maximum $M_\mathrm{gas}$ in our example for NGC 2808 older than $20$~Myr is much smaller than this, while the scale radius is also larger ($R_\mathrm{gas} =3.7$~pc). The properties of the gas cloudlets remains uncertain, but as discussed above we expect clouds that are this size and mass to be Jeans stable at temperature $200$~K. Such clouds would also be optically thick, so that obscured stars may be undetected rather than detected as reddened stars. In reality we expect some range of cloud masses and radii, but this illustrative example demonstrates that if enough material is in small clouds then this reduces the influence of extinction on the stellar population.}

\subsubsection{Constraints on the gas content}

{In terms of signatures of gas in the cluster, \citet{CabreraZiri15} inferred an upper limit of $\lesssim 9$~percent of the total cluster mass from an ALMA search for CO $J=3-2$ transitions in three young massive clusters in the Antennae, aged $50-200$~Myr. This upper limit is not strongly restrictive for the companion accretion model, since the gas reservoir (Figure~\ref{fig:ejecta_mass}) never approaches this limit. Further, in the scenario we explore in this work the heating source is the Lyman-Werner photons, which photodissociate the CO over several $100$~Myr timescales \citep{Conroy11}. Even if the CO is not fully dissociated, the $3-2$ transition may in some cases be optically thick \citep{Polychroni12, Salak14, Fukui15}, particularly if gas is concentrated into small cloudlets as discussed above.}

{Since there is little or no ongoing star formation, the fully ionised assumption adopted by \citet{Bastian13b} need not hold, such that an absence of strong \textsc{H}$\alpha$ signatures is expected. More restrictive are the searches for HI gas in 13 LMC and SMC young clusters ($30-300$~Myr old) by \citet{Bastian14}, who inferred an upper limit of $1$~percent of the total cluster mass. While this is comparable to the maximum gas mass in the model assuming rapid sweep-up accretion of the contaminants, it remains an uncomfortable constraint if accretion is slower. Once again, this constraint may be mitigated in some cases where \textsc{HI} can be optically thick \citep{Fukui15, Seifried22}. Following the discussion in Appendix B of \citet{Seifried22}, this can occur when the column density of \textsc{HI} is $\gtrsim 10^{-18}$~g~cm$^{-2}$, depending on the temperature. The gas in small clouds with mass $M_\mathrm{cloud} \sim 0.1 \, M_\odot$ and radii $R_\mathrm{cloud} \sim 10$~au would have much larger column density than this threshold. }

\subsubsection{Survival of gas against stripping }
{Finally, the clumpiness of the gas reservoir may have important consequences for the ram pressure stripping as the globular cluster moves through the external medium. \citet{Chantereau20} assumed a smooth density distribution when showing that a gas reservoir could be stripped efficiently from a globular cluster moving through a dense gaseous environment. However, for a clumpy reservoir the external, high velocity medium may pass through low density channels, similar that seen in simulations of stellar winds during the formation of stellar clusters \citep[e.g.][]{Rogers13}. This would reduce the efficiency of stripping, allowing the ejecta products to remain bound to the cluster for longer.}

{In summary, a sub-structured gas reservoir may survive long enough to be captured by the stellar population. A combination of relatively rapid accretion of the gas, photodissociation and high optical depth resulting from substructure may contribute to the non-detection of such a gas reservoir in young massive clusters. }

\subsection{Mixing of companion material with host star}
\label{sec:mixing}

\subsubsection{Binary collisional mixing}

{In order to produce the observed abundance variations, mixing of the secondary material in the majority of the host star is required.} This is in order to produce the same chemical signatures in MS and MSTO as in RGB stars \citep[e.g.][although we expect some variation in the degree of mixing -- see discussion in Section~\ref{sec:outlook_abundances}]{Briley96,Briley04}. RGB stars have convective envelopes that encompass $\sim 70$~percent of the total mass, while for MSTO stars at $10$~Gyr this is closer to $\sim 1$~percent. Hence, we must ask whether it is plausible to expect mixing of the pollutant in the scenario we consider. 

Some authors have performed smoothed particle hydrodynamics simulations of such collisions, primarily for their influence on blue straggler properties \citep[e.g.][]{Benz92, Lombardi96, Lombardi02}. It is not clear how accurate such models are for computing the degree of shock heating and convective mixing induced by a collision. Nonetheless, the studies agree that the degree of mixing is dependent on whether the encounter is `direct' (i.e. with closest approach distance $x_\mathrm{min} \rightarrow 0$) or `grazing' (i.e. with $x_\mathrm{min} \sim R_*$, for $R_*$ the stellar radius).  {In our case, the dynamical history of the companion prior to collision with the host star is driven by an accumulation of small eccentricity changes due to successive perturbations by neighbouring stars \citep{Winter22c}.} We therefore expect the collisional encounter to be grazing in nature. This case is depicted as Case W in the bottom panels of Figure 5 of \citet{Lombardi02}. {The authors define the entropic parameter $A\equiv P/\rho^\gamma$, for $P$ pressure, $\rho$ density and $\gamma$ the adiabatic index. For radius $r$ within the star, $A$ must satisfy the requirement $\mathrm{d}A/\mathrm{d}r>0$ in thermal equilibrium. The value of $A$ is significantly increased {due to shock heating} compared to the initial conditions (Figure 1 in that work) in the majority of the primary and throughout the secondary. This suggests that the contaminant brought in by the secondary may be well mixed. As discussed by \citet{Lombardi96}, the star is initially very far from thermal equilibrium, and may undergo significant convective mixing before it returns to the MS on a thermal timescale ($\sim 1{-}10$~Myr). 

{{More recently, in their hydrodynamic simulations of a merger (or extremely grazing encounter) between a brown dwarf and a solar mass star,} \citet{Cabezon22} showed that approximately $40$~percent of the brown dwarf mass stays in the outer $30$~percent in radius of the primary, indicating a mixing fraction similar those required for the companion accretion model.} We highlight that the aforementioned models, as well as those of other authors following post-collisional evolution of MS stars \citep[e.g.][]{Glebbeek08}, are not necessarily sensitive to the later stage rotational mixing of the remnant. }

We conclude that the collision of the companion with the star and the resultant merged star's subsequent evolution plausibly results in deep mixing. Unless otherwise stated, we will proceed on the assumption that sufficient mixing occurs to evenly distribute the contaminants through $70$~percent of the star, while only initiating a deviation from the MS that is much shorter than the star's life-time. 

\subsubsection{Lithium survival}

{Lithium may survive accretion of a companion with mass ratio $q\sim 0.1$ companion, as in the simulations of \citet{Lombardi02} and \citet{Cabezon22}. This may explain why there is no strong correlation in observed between other abundance variations and lithium abundances \citep[e.g.][]{Dobrovolskas14}. However, some degree of variation should be expected simply due to the different composition of the secondary, which would presumably be Li poor. One mitigating factor may be that the two individual stars of mass $m_1$ and $m_2$ are lower mass than the collisional product $m_1+m_2$, and therefore deplete their lithium more slowly than a star that has always had the mass $m_*=m_1+m_2$ \citep[e.g.][]{Bildsten97}. This is unlikely to significantly influence lithium evolution in the primary for small $q$, but may allow any residual lithium to survive on the low mass secondary. }

\subsubsection{Hertzsprung-Russell diagram}

{Another requirement for the merger is that it does not produce strong features in the Hertzsprung-Russell diagram that could be interpreted as large age disparity with unpolluted stars. This is because \citet{Martocchia18b} found that the multiple populations NGC 1978 ($\sim 2$~Gyr old) clusters are consistent with being coeval within $\sim 20$~Myr when using optical CMDs of SGB stars \citep[also in NGC 2121 -- see][]{Saracino20} or within $\sim 65$~Myr when using the MSTO width. As noted by \citet{Martocchia18b}, this finding \textit{`suggest[s] that multiple populations may be due to a stellar evolutionary effect not yet recognized in standard evolution models. This effect would need to only efficiently operate in stars within massive/dense stellar clusters'.} Interestingly, other young clusters do exhibit extended or bimodal MSTOs \citep[e.g.][]{Milone09}, although this is now generally considered to be unrelated to stellar age \citep{Cordoni22}. The breadth of these extended MSTO may be attributed to stellar rotation \citep{Bastian09, dAntona15}, which mimics the age spreads proportional to the age of the cluster \citep{Niederhofer15}. Magnetic braking would then suppress this later in the stellar life-time, explaining the lack of extended MSTOs after a few Gyr.}

{\citet{Wang20} used theoretical binary evolution models to show that the MSTO width can also be extended when massive binaries interact. This may produce an ersatz age spread that is proportional to the age of the cluster, similar to the stellar rotation hypothesis and the empirical constraints \citep{Bastian16}. If the merger of binaries in the companion accretion model would produce an extended MS in the F814W band used by \citet{Martocchia18b} to constrain age differences between populations, then this would contradict the model. However, there are two reasons why this is not necessarily the case:}
\begin{enumerate}
    \item {\citet{Wang20} studied massive binaries with MS lifetime up to $\sim 40$~Myr. In the context of the companion accretion model, we are interested in mergers with lower mass stars ($m_* \sim 1\, M_\odot$), with companions that have typical mass-ratios $q\sim 0.1{-}0.3$. Both absolute mass and companion mass-ratio are important in the evolution of rotation, magnetic fields, and H-R diagram position. The results therefore do not necessarily generalise trivially to the distributions of stellar masses and mass-ratios explored in this work.}
    \item {Mergers extend the MS in part because of their influence on the stellar rotation, which produces something similar to the observed extended MS in numerous clusters. This extension is less pronounced at longer (optical) wavelengths \citep{Brandt15}. It is unclear how the \citet{Wang20} results would appear in the F814W band used by \citet{Martocchia18b} to constrain age differences between populations. }
\end{enumerate}{We therefore consider this possibility as an important future test of the companion accretion model, but do not conclude that the results of \citet{Wang20} in conjunction with the observational results of \citet{Martocchia18b} rule out mergers at present.}

\subsection{Fraction of accreted companions}
\label{sec:rates}

We now turn our attention to quantifying the rate at which sub-stellar companions can be accreted onto the primary star. We base these estimates on the analytic prescription presented by \citet{Winter22c}, which is developed from the theoretical cross-sections for eccentricity excitation in dense clusters \citep{Heggie96}, and has been benchmarked against the numerical simulations of \citet{Hamers17}. In brief, numerous hyperbolic encounters in high velocity dispersion environments result in a diffusive eccentricity evolution for a binary system. When a high eccentricity is reached, a companion can either circularise to produce a short period companion (i.e. a hot Jupiter if $q$ is small), or it can become disrupted and/or merge with the host star. For sufficiently high density environments, rapid eccentricity fluctuations effectively forbid circularisation. For the typical semi-major axes and stellar densities we consider here, we are always in the regime where collisions are more probable than circularisation. We will therefore assume that no companions survive circularisation, but are instead accreted onto their host star. Over sufficient time, all companions will eventually either become ionised by a close dynamical perturbation or they will merge. The overall number of collisions is therefore dependent on the relative rates of these two outcomes among the population.

The rate of collision of the companion with the host star can be estimated \citep{Winter22c}:
\begin{equation}
\label{eq:Gamma_tide_main}
    \Gamma_\mathrm{coll} \approx \frac{\gamma_0  e_0 \sqrt{1-e_0^2}}{2(1-e_0)}
\end{equation} where $e_0$ is the initial eccentricity and we have defined the factor
 \begin{equation}
 \label{eq:gamma}
     \gamma_0 \equiv  4.6 \, \sqrt{1+q} \mathcal{M}_{*}^\mathrm{(coll)} \frac{n_*}{10^5 \, \rm{pc}^{-3}}  \left( \frac{m_* }{1\, M_\odot}\right)^{1/2} \,  \left(\frac{a_\mathrm{0}}{5\, \rm{au}}\right)^{3/2}  \, \rm{Gyr}^{-1}, 
 \end{equation}for
\begin{equation}
\mathcal{M}_{*}^\mathrm{(coll)}  =  \int_0^\infty \, \mathrm{d}m_\mathrm{pert} \, q_\mathrm{pert}\, \xi(m_\mathrm{pert}).
\end{equation} In the above equations, $n_*$ is the local stellar density, $\sigma_v$ is the three dimensional velocity dispersion, $a_0$ is the initial semi-major axis of the companion, $q_\mathrm{pert}= m_*/m_\mathrm{pert}$ is the mass-ratio of a perturber with mass $m_\mathrm{pert}$ and 
\begin{equation}
\xi(m_*) \propto \begin{cases}  m_*^{-\alpha_1} \qquad & m_\mathrm{min} \leq m_* <m_\mathrm{br} \\
m_*^{-\alpha_2} \qquad &  m_\mathrm{br} \leq m_*\leq m_\mathrm{max}\\ 
0 \qquad \qquad & m_*>m_\mathrm{max} \, \mathrm{or} \, m_* < m_\mathrm{min}
\end{cases}    
\end{equation}is the local mass function. We adopt $\alpha_1=0.4$, $\alpha_2=2.8$, $m_\mathrm{br} = 0.8\,M_\odot$, $m_\mathrm{min}=0.08\, M_\odot$ and truncate above $m_\mathrm{max}=1.3\, M_\odot$ to approximately replicate the mass function at $5$~Gyr. However, this choice does not strongly influence our results because the overall dominant perturbations are due to stars below this mass.

{The initial eccentricity $e_0$ in this context is not immediately obvious. We expect the binary formation via gravitational instability in a more compact disc than for hierarchical star formation. In the field, wide binary orbits are generally not circular, but exhibit a wide range of eccentricities \citep[e.g.][]{Duqennoy91}. In fact, binaries at separations $\lesssim 100$~au, that may have formed through gravitational instability in a disc, exhibit an approximately uniform distribution of eccentricities \citet{Hwang22}. We will therefore adopt $e_0=0.5$ as the mean of the uniform eccentricity distribution. However, we highlight that if eccentricities are small then equation~\ref{eq:Gamma_tide_main} implies the collision rate is also small (and vanishes as $e_0\rightarrow 0$). This is because the dominant terms in the dynamical cross sections vanish at $e_0=0$, such that only close encounters play an important role in the initial changes to eccentricity evolution \citep{Ostriker94, Winter18}. In this case, the collision rate may be initially slower, but the early eccentricity excitation must be treated with an alternative prescription that factors in higher order terms. }

From equation~\ref{eq:gamma}, we see that the rate of enrichment by collision with the companion is independent of the local velocity dispersion, but linearly dependent on the local density.  This approximation does not necessarily apply when the orbital {frequency} of the companion becomes comparable to the encounter rate \citep{Kaib14}, or more stringently becomes less accurate when the majority of encounters are not `slow' (i.e. when $\sigma_\mathrm{v} \gg  v_\mathrm{orb}$, for the orbital velocity $v_\mathrm{orb} \sim 10$~km~s$^{-1}$; see discussion in Appendix A of \citealt{Winter22c}). However, since $\sigma_v$ is never more than a factor few larger than the orbital velocity, we are satisfied that our estimate should be a reasonable approximation. 

By comparison, the rate at which a star-companion system undergoes a resonant or ionising encounter can be approximated \citep{Hut83}:
\begin{multline}
\label{eq:Gamma_ion}
    \Gamma_\mathrm{ion} = 2.8 \mathcal{M}^{\rm{(ion)}}_* \left(\frac{ \sigma_v}{10 \,\mathrm{km\,s}^{-1}}\right)^{-1}\frac{m_*}{1\, M_\odot} \times \\ \times \frac{ a_0}{5 \, \rm{au}} \frac{n_*}{10^5 \, \rm{pc}^{-3}}  \, \mathrm{Gyr}^{-1}.
\end{multline}Here we have defined:
\begin{equation}
    \mathcal{M}^{\rm{(ion)}}_*  = \int_0^\infty \mathrm{d}m_\mathrm{pert}\,  (1+q_\mathrm{pert}+q) q_\mathrm{pert}^{1/3}  \xi(m_\mathrm{pert}).
\end{equation}We consider all strong dynamical perturbations of this kind to be ionising as a necessary simplification. Strong encounters may in reality alter the energy of the companion, which then may subsequently be ionised or accreted as before (albeit with a different semi-major axis). Equation~\ref{eq:Gamma_ion} therefore may moderately overestimate the role of ionisation relative to collisions. However, we immediately infer from a comparison between equation~\ref{eq:Gamma_ion} and equation~\ref{eq:gamma} that for sufficiently large velocity dispersions, accretion of the companion dominates over ionisation. 

With these expressions, we can estimate the relative probability of a star having experienced enrichment by the accretion of the companion at time $t$:
\begin{equation}
\label{eq:Penr}
    P_\mathrm{coll} = \frac{\Gamma_\mathrm{coll}}{\Gamma_\mathrm{tot}} \left\{ 1- \exp\left[-\Gamma_\mathrm{tot} t \right] \right\},
\end{equation}where $\Gamma_\mathrm{tot} = \Gamma_\mathrm{coll}+\Gamma_\mathrm{ion}$. In the absence of any further dynamical effects (see Section~\ref{sec:depletion}), this simple equation gives the overall fraction of stars that are contaminated with pollutants. To first order, at $t\gg 1/\Gamma_\mathrm{tot}$, we see that $P_\mathrm{coll} \propto \sigma_v \propto \sqrt{M_\mathrm{c} /R_\mathrm{c}}$ for total cluster mass $M_\mathrm{c}$ and radius $R_\mathrm{c}$. Given the lack of clear correlation between $M_\mathrm{c}$ and $R_\mathrm{c}$ among globular clusters \citep{Krumholz19}, this suggests $P_\mathrm{coll} \propto M_\mathrm{c}^{1/2}$. This positive correlation is in broad agreement with what is observed \citep[see][and references therein]{Bastian18}, however we return to this in more detail in Section~\ref{sec:obs_comp}. 

When we compute the rate of capture and ionisation we will generally adopt the central properties of the cluster within the scale radius $a_\mathrm{c} = R_\mathrm{hm}/1.305$ in a Plummer sphere. We choose the central values because this is where the most rapid encounters occur, and the velocity dispersion in this region is therefore the most important factor in setting the ratio of enrichment to ionisation. Indeed, {if stars outside of the core are more likely to pass beyond the tidal radius and be lost from the cluster, then} the stars that survive to the present day should also be biased to those that occupied the cluster core earlier in the cluster evolution. Meanwhile, the assumption of a Plummer sphere may not always be the most accurate profile, however it allows us to homogeneously compare across different clusters with simple scaling relations while adopting the half-mass radius from observations. The latter is a comparatively robust quantity, that does not depend on density profile definitions and should not vary rapidly with cluster evolution. It is probable that deviations in the present day true central density compared to this assumption are less important than the variations over the dynamical history of a given cluster. We leave these considerations of more accurate dynamical modelling to future work.

\subsection{Removal of pristine stars}
\label{sec:depletion}

 \begin{figure}
    \centering
    \includegraphics[width=\columnwidth]{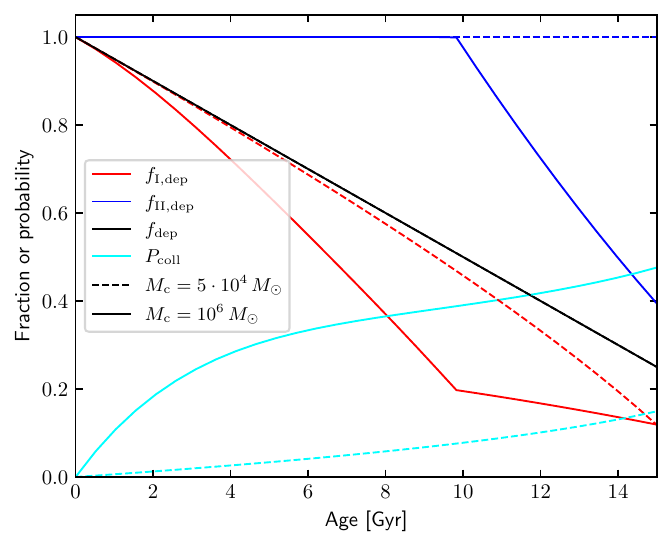}
    \caption{The assumed scaling of the depletion factors for a fixed cluster mass $M_\mathrm{c} = 10^6 \, M_\odot$ (solid lines) and $5\times 10^4 \, M_\odot$ (dashed lines) over time. The red lines show the depletion of the non-polluted population of stars, while the blue lines show the equivalent for the polluted stars. The black lines show the overall assumed depletion factor $f_\mathrm{dep} = M_\mathrm{c}/M_\mathrm{c,0}$, and the cyan lines shows the probability of a star having been enriched due to accretion of the sub-stellar companion. All clusters are assumed to have a current half-mass radius $R_\mathrm{hm}=10$~pc. Note, this figure does not show the evolution of a specific cluster, but the associated fractions for a fixed mass at the specified age (the initial mass $M_\mathrm{c,0} = M_\mathrm{c}/f_\mathrm{dep}$ increases with decreasing $f_\mathrm{dep}$).}
    \label{fig:fdep_evol}
\end{figure}

The removal of the pristine population over time is a necessary ingredient of all enrichment models to enhance the overall enrichment fraction. The loss of some fraction of the total mass of a cluster over time is naturally expected due to the two-body relaxation of the cluster, as well as by other sources of dynamical heating such as tidal shocks. {In models that invoke a second generation of stars (not companions),} a second population may be expected to form in the core region of the cluster \citep[e.g.][]{DErcole08, Calura19}, so that the assumption that the pristine population is preferentially removed is to some extent justified. However, the fraction of the pristine population that must be removed in multi-generation models is usually extremely high ($>90$~percent), unless some mechanism for significant dilution can be invoked \citep[e.g.][]{Conroy11}. 

In the context of the mechanism we present here, we also naturally expect the contaminated population to survive in much higher numbers than the pristine population. {This is partly because the stars in the core, where accretion of the companion is most probable due to higher density and $\sigma_v$, are also the most gravitationally bound. In addition, we might expect some correlation between stars that are ejected and those that remain pristine through dynamical arguments.} The energy input required to ionise the star-companion system is:
\begin{equation}
    \Delta E_\mathrm{ion} = \frac{G m_*^2 q}{a_0}.
\end{equation}This is generally smaller than the energy input required to unbind the star from the cluster:
\begin{equation}
\label{eq:Ebind}
    \Delta E_\mathrm{unbind} = - E_\mathrm{i} \sim  -\frac{1}{2}m_* v_*^2 +\frac{G M_\mathrm{c} m_* }{a_\mathrm{c}} \left(1+\frac{r_*^2}{a_\mathrm{c}^2}\right)^{-1/2},
\end{equation}where $E_\mathrm{i}$ is the initial energy of the stellar orbit in the cluster and the RHS of equation~\ref{eq:Ebind} is an approximation adopting a Plummer density profile with scale radius $a_\mathrm{c}$, where a star is at radius $r_*$ from the centre and moving with velocity $v_*$. For the sake of illustration, plugging in $v_* = 10$~km~s$^{-1}$, $M_\mathrm{c} = 10^5 \, M_\odot$, $r_* = a_\mathrm{c} = 2$~pc, $a_0= 5$~au, $m_* = 1\, M_\odot$ and $q=0.1$, we obtain $\Delta E_\mathrm{unbind}/\Delta E_\mathrm{ion} = 5.75$. {Thus only a small fraction of an encounter that unbinds the star needs to go into the binary orbit in order to result in ionisation. However, this is only the case if stars become unbound due to individual encounters that impart large kinetic energy changes, and it is unclear whether this is the case in practice. Preferential loss from the cluster of binaries that have been ionised would lead to an increased binary fraction in the cluster core. However, it is not straight forward to interpret this from simulations that consider evolving binaries where $q\sim 1$ \citep[e.g.][]{Hong19}, since other effects such as mass segregation also play a role. We will here assume preferential loss of the pristine population, but the degree to which this is true needs to be investigated further in future simulations.}

The consequences of a depletion in stellar mass by a factor $f_\mathrm{dep} \equiv M_\mathrm{c}/M_\mathrm{c,0}$ for the overall fraction of polluted stars in the companion accretion model is two-fold. {In the first instance, the fact that a cluster has been depleted means that its initial mass was larger. We will generally assume the half-mass radius does not change substantially. Thus the increase in mass yields an increase in initial density and velocity dispersion, somewhat altering the pollution fraction via the local density and relative ionisation and collision rates (see Section~\ref{sec:rates}). More importantly, the depletion can reduce the number of pristine stars remaining in the cluster. The total depletion factor can be written:
\begin{equation}
\label{eq:fdep}
    f_\mathrm{dep} = (1-P_\mathrm{coll})f_\mathrm{I,dep} + P_\mathrm{coll} f_\mathrm{II,dep}
\end{equation}where $f_\mathrm{I,dep}$ and $f_\mathrm{II,dep}$ are the depletion factors for the population I (pristine) and population II (polluted) stars respectively. With these definitions, we can determine 
\begin{equation}
\label{eq:fI}
   f_\mathrm{I} =   \frac{f_\mathrm{I,dep} (1-P_\mathrm{coll})}{f_\mathrm{I,dep} (1-P_\mathrm{coll}) + f_\mathrm{II,dep} P_\mathrm{coll}},
\end{equation}which is the fraction of unpolluted stars remaining in the cluster. We discuss sensible functional forms of $f_\mathrm{I,dep}$ and $f_\mathrm{II,dep}$ in Appendix~\ref{app:reldep}. We expect differential depletion between the two populations because the collisions occur preferentially when a star is in the core of the cluster. This is where the density is largest and the timescale for encounters is shortest. If $P_\mathrm{coll}$ (the collision probability) is low, then this should mean nearly all collisions have happened in the core, where they are most frequent. These stars should also therefore be those should take the longest to be ejected by two-body encounters. On the other hand, as $P_\mathrm{coll}$ grows, eventually stars that have undergone collision are distributed throughout the cluster. This is because there is a relatively large chance of having had a collision even for stars outside the centre. }

Examples of the assumed scaling for the depletion factors of the polluted and pristine populations are shown in Figure~\ref{fig:fdep_evol}. To generate the depletion factors in a given cluster as a function of time, we have assumed that:
\begin{equation}
\label{eq:fdep_2}
    f_\mathrm{dep} = 1 + (f_\mathrm{dep,0}-1) \frac{t}{t_0},
\end{equation}with $f_\mathrm{dep,0}=0.5$ at time $t=t_0=10\,$Gyr. This corresponds to a simple, linear decrease in the stellar mass that very approximately mimics the influence of two body encounters over time. The normalisation is chosen such that $f_\mathrm{dep} \sim 0.3{-} 0.5$ at the typical ages of globular clusters \citep[e.g.][]{Kruijssen15, Webb15}. For the two different fixed cluster masses are adopted at each time, $M_\mathrm{c} =10^6\, M_\odot$ and $M_\mathrm{c} =5\cdot 10^4\, M_\odot$, both with half-mass radii $R_\mathrm{hm}$. The two regimes of the depletion behaviour for each population can be seen at later times in the higher mass case. {As discussed above, the depletion of the polluted (II) stars becomes less efficient relative to the pristine (I) stars when a large fraction of the stars in the cluster, including those outside of the core, become polluted. In the lower mass model, $P_\mathrm{coll}$ stays low for all ages, meaning that the polluted stars are preferentially in the core. The polluted population is therefore never large enough to become depleted in this case. We emphasise that this is not a unique solution, but is a physically motivated toy model as discussed in Appendix~\ref{app:reldep}.} Broadly, our toy depletion model reproduces the kind of factors we might expect, and completes the model to be compared to observations in Section~\ref{sec:obs_comp}.

\subsection{Trends with globular cluster properties}
 \label{sec:obs_comp}
 
 In the following we will consider how our models reproduce the observational trends in the enrichment fractions in various globular clusters. For simplicity, we will always assume that the initial semi-major axis of the companion has $a_0 =5$~au (approximately half the BHL radius) and the companion eccentricity is initially $e_0=0.5$, the average thermalised value. All other assumptions and calculations are as outlined in the preceding section. Although our models are highly simplified, we have deliberately kept our model simple and fixed parameters in order to avoid fine-tuning of the model to the data. Our concern is mostly on the trends predicted by the model rather than the exact fit of the data, although the parameters choices are all physically or empirically motivated.

 \subsubsection{Presence of multiple populations}
  \begin{figure*}
    \centering
    \includegraphics[width=0.8\textwidth]{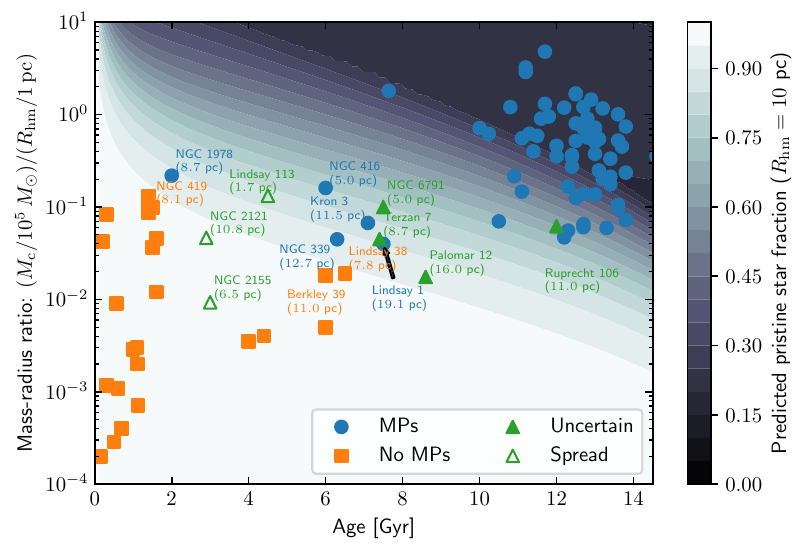}
    \caption{The distribution of observed clusters that have been searched for evidence of multiple populations (MPs), as compiled by \citet{Krause16}. We include only globular clusters with age estimates in that dataset. We have added a number of clusters from the recent works of \citep{Martocchia17, Martocchia18a,Martocchia19, Saracino20} and Palomar 12, Terzan 7, NGC 6791 and Ruprecht 106 are designated `uncertain', mostly due to the low sample sizes \citep[see][and references therein]{Bastian18}. {From \citet{Martocchia19}, we have denoted Lindsay 113, NGC 2121 and NGC 2155 with empty triangles. These clusters exhibit a larger spread in abundances than expected from observational uncertainties, but Gaussian mixture models do not demonstrate the presence of discrete populations.} The colour bar is a guide as to the enrichment fraction with an assumed present day half-mass radius $R_\mathrm{hm}=10$~pc and a depletion factor $f_\mathrm{dep}$ that varies such that $(1-f_\mathrm{dep}) = 0.5 \cdot  10\,\rm{Gyr}/t$ for age $t$. The names of some notable clusters are indicated, with half-mass radii shown in brackets. {Note that where the half-mass radius $R_\mathrm{hm}<10$~pc this can result in reduced pristine fractions at early times with respect to the colourbar estimate (and \textit{vice versa}, see text for details).}}
    \label{fig:MP_YN}
\end{figure*}

 In our statistical comparison between the model we have presented in the preceding section and the rate of stellar enrichment in globular clusters, we first consider the regions for which evidence of multiple populations has been found compared with those that have not. Within the companion accretion model, the age and velocity dispersion are the two most important parameters in determining the fraction of polluted stars. {The virial velocity dispersion scales with $\sqrt{M_\mathrm{c}/R_\mathrm{hm}}$, and we therefore consider globular clusters in terms of ${M_\mathrm{c}/R_\mathrm{hm}}$ and age.} In our model, we assume that the total cluster population depletion follows equation~\ref{eq:fdep_2}, with $f_\mathrm{dep} = 0.5$ at $t=10$~Gyr. 
 
 In Figure~\ref{fig:MP_YN} we show the distribution of cluster properties divided by whether or not they show evidence of harboring multiple populations \citep[][]{Krause16}. {We have further included/modified the clusters surveyed by \citet{Martocchia18a} and \citet{Martocchia19}, with cluster parameters estimated by \citet{Song21} except for Lindsay 113 and Lindsay 38 for which we use an approximate half-mass radius based on the findings of \citet{Bica00}.} When compared with the predicted fraction of pristine (unpolluted) stars in the model {(using equation~\ref{eq:Penr} combined with equation~\ref{eq:fI})}, we find generally good agreement in terms of {clusters in which} we expect to find multiple populations. {In particular, we predict a non-negligble polluted population in the young NGC 1978, which is only $2$~Gyr old and has two, coeval populations \citep{Martocchia18a, Saracino20}. Using the actual properties of the cluster in our estimate ($M_\mathrm{c} = 1.9\times 10^5 \, M_\odot$, $R_\mathrm{hm} = 8.7$~pc and $t=1.9$~Gyr), we expect a moderate fraction ($f_\mathrm{II} = 5$~percent assuming $f_\mathrm{dep}=1$, or $f_\mathrm{II} = 21$~percent if $f_\mathrm{dep}=0.5$) of polluted stars. Based on Figure 5 in the study \citet{Martocchia18a}, this fraction seems reasonable. This polluted population can therefore be explained by our model.}
 
 {In three instances, NGC 2121, NGC 2155 and Lindsay 113, \citet{Martocchia19} do not find evidence of two populations in their Gaussian mixture models, but do find N spreads that are broader than can be accounted for by observational error alone (empty green triangles in Figure~\ref{fig:MP_YN}). Companion accretion may feasibly contribute to this spread. For example, in the case of NGC 2155, the half-mass radius is $R_\mathrm{hm}\approx 6.5$~pc \citep{Song21}, which means that the cluster is higher density than our fiducial estimate. Therefore both ionisations and collisions occur more rapidly. When substituting $R_\mathrm{hm}= 6.5$~pc, $M_\mathrm{c} = 6 \times 10^3 \, M_\odot$, $t=3$~Gyr into our model, we obtain $f_\mathrm{II} = 0.01$ for $f_\mathrm{dep}=1$ or $f_\mathrm{II}=0.03$ for $f_\mathrm{dep}=0.5$, such that a small fraction of stars may have been enriched by companion accretion. Similarly, a few percent of the stars in NGC 2121 may be polluted via companion accretion. }
 
 The marginal cases where no conclusive evidence of multiple populations have been found among small sample sizes, Terzan 7 and Palomar 12, sit close to the border where a non-negligible fraction of polluted stars are expected. The most challenging case is that of Ruprecht 106. In this case, 9 stars have been searched for evidence of Na-O abundance variations \citep{Villanova13}. Assuming $f_\mathrm{dep}=0.5$, our model with the properties of Ruprecht 106 ($R_\mathrm{hm}=11$~pc, $M_\mathrm{c} = 8.3\times 10^4 \, M_\odot$, $t=12$~Gyr) {predicts a $82$~percent pristine fraction, which gives a $17$~percent probability of finding $9$ pristine stars randomly, which is not statistically significant. Further, for a small amount of depletion ($f_\mathrm{dep} =1$), this increases to  $95$~percent pristine fraction ($63$~percent chance).} 
 
{{There remains a curious case in the cluster NGC 6791, which has a mass of} $M_\mathrm{c} \approx 5\times 10^4 \, M_\odot$ and radius $R_\mathrm{hm} \approx 5 $~pc and age $t\approx 7.5$~Gyr, for which our model would yield $f_\mathrm{I} = 88$~percent if $f_\mathrm{dep}=1$ and $67$~percent if $f_\mathrm{dep}=0.5$. As dicussed by \citet{Villanova18}, the nature of NGC 6791 is a controversial topic, it being an old cluster with a very high metallicity \citep[][and references therein]{Cunha15}. Different authors have characterised it both as an open cluster, globular cluster, or remnant dwarf galaxy and a member of the thick disc, thin disc, or the galactic bulge. While \citet{Geisler12} found evidence of inhomogeneities and a Na-O anti-correlation, subsequent studies have not reproduced this finding. In particular,  \citet{Villanova18} did not find evidence of inhomogeneities, despite choosing the same sample as \citet{Geisler12}. This was a sample of $17$~stars, which with $f_\mathrm{dep}=1$ would give a $11$~percent chance of a null result, but $0.1$~percent if $f_\mathrm{dep}=0.5$. Given the peculiarity of this cluster, we do not consider this a strong counterexample for our model. We conclude that the predictions from the companion accretion model for the conditions under which multiple populations are found remain largely consistent with current constraints.}

{Finally, we highlight that all of the studies of young clusters shown as squares in Figure~\ref{fig:MP_YN} were performed on post-main sequence stars. By contrast, \citet{Cadelano22} find hints of a N spread in the main sequence stars of the $1.5$~Gyr old NGC 1783. The absence of this signature for evolved stars may be due to first dredge up mixing wiping out apparent N differences between their surface abundances \citep{Salaris20}. In the context of the companion accretion model, some abundance spread among young MS stars may result from surface contamination in a similar manner to the early disc accretion scenario \citep{Bastian13}.}

 \subsubsection{Scaling with cluster mass}

 \begin{figure*}
    \centering
    \includegraphics[width=0.8\textwidth]{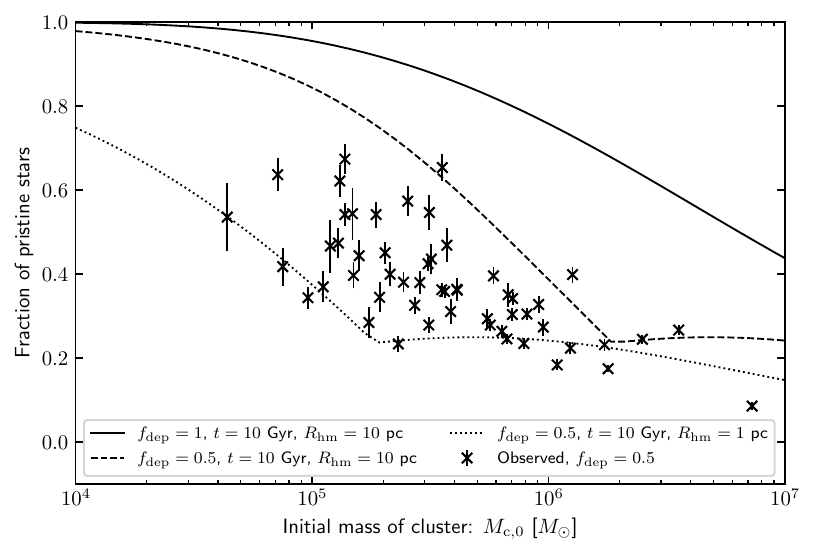}
    \caption{The number of unpolluted stars based on the compilation of \citet{Milone17}, shown as crosses with assumed depletion factor of $f_\mathrm{dep}=0.5$ to yield the initial masses. The lines represent three different illustrative examples for the outcome of the enrichment model put forward in this work. The solid line includes no depletion $f_\mathrm{dep}=1$, with an assumed age of $10$~Gyr and cluster half-mass $R_\mathrm{hm}= 10$~pc. The dashed line is the same, but with a more reasonable depletion factor $f_\mathrm{dep}=0.5$. The dotted line has the same depletion factor $f_\mathrm{dep}=0.5$ and a smaller  $R_\mathrm{hm}= 1$~pc. All models assume an initial semi-major axis for the companion of $a_0=5$~au. }
    \label{fig:fenrich_Mc}
\end{figure*}

We will now consider how the enrichment fraction correlates with the (initial) cluster mass. For this purpose, we use the enrichment fractions compiled by \citet{Milone17}. Since the globular clusters are all old, we here assume that they have all have the same depletion factor $f_\mathrm{dep} = 0.5$ rather than compounding uncertainties in the dynamical timescales, tidal histories and ages to estimate the depletion factor for each cluster individually. 

We show these data in Figure~\ref{fig:fenrich_Mc}, where we also show the expected fraction of pristine (unpolluted) stars expected from our models, assuming an age $t=10$~Gyr and either a depletion factor $f_\mathrm{dep}=1$ or $0.5$ and a half-mass radius $R_\mathrm{hm}= 10$~pc or $1$~pc. We do not introduce a mass-radius relation since any such trends for globular clusters is weak, washed out by scatter \citep[see Figure 9 of][]{Krumholz19}. We find that if no depletion is considered then our model still predicts a similar trend with the initial cluster mass, however it overestimates the number of unpolluted stars remaining. When we adopt a modest $f_\mathrm{dep}=0.5$, the fraction of unpolluted stars decreases due to the preferential removal of these stars. These models are in quantitative agreement with observed pristine fraction in \citet{Milone17} dataset.  Qualitatively, the fraction of pristine stars exhibits a `knee' at the transition from the assumption that all stars that are dynamically ejected are pristine, to a mix of the two (see Appendix~\ref{app:reldep}). This naturally produces a greater spread in the fraction of polluted stars at lower cluster masses, similar to what is found empirically. Quantitatively and qualitatively, our model is a good description of the enrichment trends with cluster mass. 
 
 \subsubsection{Individual cluster enrichment fractions}

\begin{figure*}
    \centering
    \includegraphics[width=0.8\textwidth]{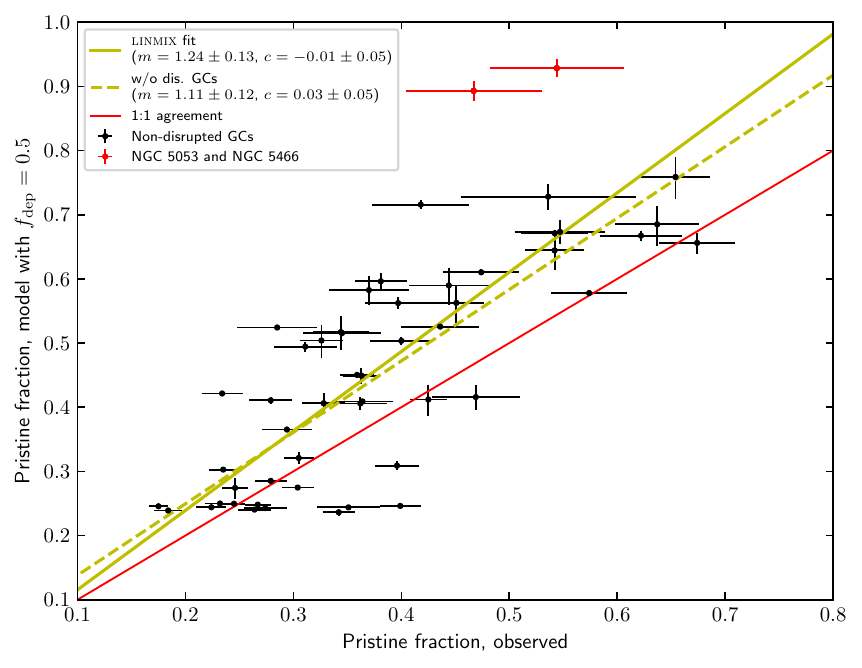}
    \caption{The number of unpolluted stars in globular clusters as compiled by \citet{Milone17}, compared with the prediction of our model using the dynamical data of \citet{Hilker20} with a depletion factor $f_\mathrm{dep}=0.5$ and initial semi-major axes $a_0 = 5$~au. The uncertainties in the y-axis (model) are obtained using the uncertainty in the total cluster mass in the database by \citet{Hilker20}. 
    The red line is the $1:1$ line, along which the model agrees completely with the data. The yellow lines have been obtained using the \textsc{linmix}\protect\footnotemark fitting procedure \citep{Kelly07}, the solid line including all datapoints while the dashed line excludes the two red points that correspond to the tidally disrupted clusters NGC 5053 and NGC 5466 \citep{Belokurov06, Lauchner06}.}
    \label{fig:obs_vs_model}
\end{figure*}
\footnotetext{\url{https://github.com/jmeyers314/linmix}}

We now consider how well the model we have put forward reproduces the enrichment fraction for each cluster based on their specific dynamical properties. Specifically, we take the masses and half-mass radii of the globular clusters in the compilation by \citet{Milone17} from the database of \citet{Hilker20}, and the ages as compiled by \citep{Krause16}. We then use our model to predict, for a fixed depletion factor $f_\mathrm{dep}=0.5$, what the observed fraction of unpolluted stars should be. 

The outcome of this exercise is shown in Figure~\ref{fig:obs_vs_model}. Immediately, we find a clear correlation between the predicted and observed pristine star fraction (Spearmann rank correlation $p$-value $<10^{-11}$). To better quantify the agreement, we perform a fitting procedure using the \textsc{linmix} package \citep{Kelly07}. The best fit relationship between the model predictions $f_\mathrm{model}$ and the observed fractions $f_\mathrm{obs}$ is assumed to take the form:
\begin{equation}
    f_\mathrm{model} = m f_\mathrm{obs} +c,
\end{equation}where $m=1$ and $c=0$ correspond to perfect agreement. We obtain $m=1.24\pm 0.13$ and $c=-0.01\pm0.05$, which is remarkably good agreement given the simplicity of the model. Further excluding the clusters NGC 5053 and NGC 5466 that have both undergone tidal disruption, as evident from their extended tidal tails \citep{Lauchner06, Belokurov06}, we obtain an even better fit with $m=1.11\pm 0.12$ and $c=0.03\pm 0.05$. The minor tension between model and observations could originate from variations in the depletion rates, dynamical timescale, or initial semi-major axes (for example, via variations in the temperature of the polluted medium; see Section~\ref{sec:Bondi}) among other possibilities.

We can also turn our model around to ask `what is the depletion factor needed to obtain the observed enrichment rate?' In order to do this, we apply the MCMC implementation \textsc{emcee}\footnote{\url{https://github.com/dfm/emcee}} \citet{Foreman-Mackey13}. We adopt a log-likelihood:
\begin{equation}
    \ln \mathcal{L} = -\frac{(f_\mathrm{I, mod} - f_\mathrm{I,obs})^2}{2\sigma_\mathrm{err}^2},
\end{equation}where $f_\mathrm{I, mod}$ and $f_\mathrm{I, obs}$ are the model and observed pristine fractions respectively and $\sigma_\mathrm{err}^2 = \sigma_\mathrm{mod}^2+ \sigma_\mathrm{obs}^2 $ is the combined uncertainty. We adopt a uniform prior between $10^{-3} < f_\mathrm{dep} < 1$, and the resulting estimates are listed in Table~\ref{tab:1}. We find that for the majority of clusters $f_\mathrm{dep} \sim 0.2{-}0.6$ are sufficient to reproduce the observed fractions, with some exceptions. 

In some cases, much smaller $f_\mathrm{dep}\sim 0.01$ are required under our assumed model in order to reproduce the observed {pristine star fractions}. This may be due to extreme depletion in these cases, however we also highlight that our model struggles to produce the lowest pristine fractions $f_\mathrm{I} \lesssim 0.25$ due to assumptions in our depletion model. In particular, for small $f_\mathrm{I}$ the depleted population transitions from being $100$~percent unpolluted stars ($f_\mathrm{I,dep} = f_\mathrm{I,min}^{(1)}$, equation~\ref{eq:fI_min}) to also including enriched stars ($f_\mathrm{I,dep} = f_\mathrm{I,min}^{(2)}$, equation~\ref{eq:fI_min2}). By instead setting $f_\mathrm{I,min}^{(2)}=0$, we can obtain the maximal depletion of the pristine population to yield the factor $f_\mathrm{dep,max}$ (i.e. no polluted stars are lost until no pristine stars are left). Once we do this, we obtain the results shown again in Table~\ref{tab:1}, which now show practically all clusters with $f_\mathrm{dep,max}>0.25$, consistent with inferred depletion \citep{Kruijssen15, Webb15}. The two exceptions are the disrupted NGC 5053 and NGC 5466 \citep{Lauchner06, Belokurov06}. We therefore conclude that the depletion factors required to reproduce the observed fraction of unpolluted stars are consistent with empirical constraints. 

\begin{table*}
  \centering
  \begin{tabular}{ccccccccc}
Cluster & Mass & Half-mass radius & Tidal radius & Age & Unpol. frac. (obs) & Unpol. $f_\mathrm{dep} = 0.5$ &  Dep. factor & Maximum dep. factor  \\ 
 & $ M_\mathrm{c}$ [$10^5\,M_\odot$] & $R_\mathrm{hm}$ [pc] & $R_\mathrm{t}$ [pc] &  $t$ [Gyr] & $f_{\mathrm{I,obs}}$ & $f_{\mathrm{I,mod}}$ &  $f_{\mathrm{dep}}$ & $f_{\mathrm{dep,max}}$ \\
\hline
\rowcolor{Gray}
NGC 104 & $8.95 \pm 0.06$ & $6.30$  & $126.80$ & $12.80$ &  $0.175\pm0.009 $ & $0.25$ &  $0.046^{+0.010}_{-0.008}$ & $0.589^{+0.005}_{-0.005}$ \\
NGC 288 & $0.93 \pm 0.03$ & $8.37$  & $94.86$ & $12.20$ &  $0.542\pm0.031 $ & $0.59$ &  $0.063^{+0.413}_{-0.025}$ & $0.470^{+0.024}_{-0.023}$ \\
\rowcolor{Gray}
NGC 362 & $2.84 \pm 0.04$ & $3.79$  & $91.96$ & $10.00$ &  $0.279\pm0.015 $ & $0.24$ &  $0.445^{+0.127}_{-0.197}$ & $0.570^{+0.010}_{-0.010}$ \\
NGC 1261 & $1.82 \pm 0.03$ & $5.23$  & $140.39$ & $10.24$ &  $0.359\pm0.016 $ & $0.33$ &  $0.516^{+0.012}_{-0.347}$ & $0.519^{+0.011}_{-0.010}$ \\
\rowcolor{Gray}
NGC 1851 & $3.18 \pm 0.04$ & $2.90$  & $127.09$ & $7.64$ &  $0.264\pm0.015 $ & $0.25$ &  $0.413^{+0.143}_{-0.147}$ & $0.611^{+0.011}_{-0.010}$ \\
NGC 2298 & $0.56 \pm 0.08$ & $3.31$  & $74.71$ & $12.40$ &  $0.370\pm0.037 $ & $0.48$ &  $0.439^{+0.035}_{-0.035}$ & $0.435^{+0.038}_{-0.034}$ \\
\rowcolor{Gray}
NGC 2808 & $8.64 \pm 0.06$ & $3.89$  & $165.91$ & $11.20$ &  $0.232\pm0.014 $ & $0.25$ &  $0.380^{+0.195}_{-0.140}$ & $0.681^{+0.011}_{-0.010}$ \\
NGC 3201 & $1.60 \pm 0.03$ & $6.78$  & $77.53$ & $11.10$ &  $0.436\pm0.036 $ & $0.41$ &  $0.513^{+0.029}_{-0.036}$ & $0.517^{+0.026}_{-0.024}$ \\
\rowcolor{Gray}
NGC 4590 & $1.22 \pm 0.09$ & $7.58$  & $77.43$ & $12.70$ &  $0.381\pm0.024 $ & $0.49$ &  $0.434^{+0.022}_{-0.021}$ & $0.432^{+0.023}_{-0.022}$ \\
NGC 4833 & $2.06 \pm 0.10$ & $4.76$  & $80.15$ & $12.50$ &  $0.362\pm0.025 $ & $0.28$ &  $0.549^{+0.022}_{-0.028}$ & $0.552^{+0.022}_{-0.021}$ \\
\rowcolor{Gray}
NGC 5024 & $4.55 \pm 0.32$ & $10.18$  & $177.72$ & $12.70$ &  $0.328\pm0.020 $ & $0.28$ &  $0.529^{+0.022}_{-0.022}$ & $0.529^{+0.023}_{-0.022}$ \\
NGC 5053 & $0.74 \pm 0.16$ & $17.31$  & $97.05$ & $12.30$ &  $0.544\pm0.062 $ & $0.91$ &  $0.218^{+0.042}_{-0.037}$ & $0.209^{+0.046}_{-0.047}$ \\
\rowcolor{Gray}
NGC 5272 & $4.06 \pm 0.17$ & $6.34$  & $127.28$ & $11.40$ &  $0.305\pm0.014 $ & $0.24$ &  $0.565^{+0.014}_{-0.014}$ & $0.566^{+0.014}_{-0.014}$ \\
NGC 5286 & $3.53 \pm 0.12$ & $3.79$  & $93.89$ & $12.50$ &  $0.342\pm0.015 $ & $0.25$ &  $0.643^{+0.015}_{-0.015}$ & $0.642^{+0.016}_{-0.015}$ \\
\rowcolor{Gray}
NGC 5466 & $0.60 \pm 0.10$ & $14.03$  & $79.88$ & $13.60$ &  $0.467\pm0.063 $ & $0.87$ &  $0.238^{+0.035}_{-0.030}$ & $0.237^{+0.039}_{-0.039}$ \\
NGC 5897 & $1.57 \pm 0.14$ & $10.99$  & $69.58$ & $12.30$ &  $0.547\pm0.042 $ & $0.59$ &  $0.473^{+0.045}_{-0.041}$ & $0.474^{+0.044}_{-0.040}$ \\
\rowcolor{Gray}
NGC 5904 & $3.94 \pm 0.06$ & $5.68$  & $81.73$ & $11.50$ &  $0.235\pm0.013 $ & $0.24$ &  $0.322^{+0.171}_{-0.201}$ & $0.535^{+0.008}_{-0.008}$ \\
NGC 5986 & $3.34 \pm 0.23$ & $4.25$  & $65.86$ & $12.20$ &  $0.246\pm0.012 $ & $0.25$ &  $0.353^{+0.160}_{-0.174}$ & $0.556^{+0.019}_{-0.020}$ \\
\rowcolor{Gray}
NGC 6093 & $3.38 \pm 0.09$ & $2.62$  & $61.98$ & $12.50$ &  $0.351\pm0.029 $ & $0.25$ &  $0.697^{+0.028}_{-0.026}$ & $0.698^{+0.028}_{-0.025}$ \\
NGC 6101 & $1.78 \pm 0.32$ & $14.06$  & $89.11$ & $12.50$ &  $0.654\pm0.032 $ & $0.70$ &  $0.469^{+0.079}_{-0.074}$ & $0.462^{+0.076}_{-0.074}$ \\
\rowcolor{Gray}
NGC 6121 & $0.87 \pm 0.01$ & $3.69$  & $53.43$ & $13.10$ &  $0.285\pm0.037 $ & $0.41$ &  $0.209^{+0.215}_{-0.125}$ & $0.435^{+0.018}_{-0.016}$ \\
NGC 6144 & $0.79 \pm 0.14$ & $4.91$  & $28.18$ & $13.80$ &  $0.444\pm0.037 $ & $0.49$ &  $0.475^{+0.049}_{-0.047}$ & $0.471^{+0.051}_{-0.049}$ \\
\rowcolor{Gray}
NGC 6171 & $0.75 \pm 0.04$ & $3.94$  & $36.17$ & $13.40$ &  $0.397\pm0.031 $ & $0.45$ &  $0.462^{+0.026}_{-0.036}$ & $0.467^{+0.023}_{-0.022}$ \\
NGC 6205 & $5.45 \pm 0.21$ & $5.26$  & $134.25$ & $11.70$ &  $0.184\pm0.013 $ & $0.25$ &  $0.044^{+0.016}_{-0.011}$ & $0.557^{+0.011}_{-0.012}$ \\
\rowcolor{Gray}
NGC 6218 & $1.07 \pm 0.03$ & $4.05$  & $45.27$ & $13.40$ &  $0.400\pm0.029 $ & $0.39$ &  $0.508^{+0.022}_{-0.022}$ & $0.510^{+0.020}_{-0.020}$ \\
NGC 6254 & $2.05 \pm 0.04$ & $4.81$  & $54.06$ & $12.40$ &  $0.364\pm0.028 $ & $0.28$ &  $0.550^{+0.021}_{-0.019}$ & $0.550^{+0.020}_{-0.019}$ \\
\rowcolor{Gray}
NGC 6341 & $3.52 \pm 0.04$ & $4.49$  & $120.87$ & $13.20$ &  $0.304\pm0.015 $ & $0.25$ &  $0.592^{+0.011}_{-0.011}$ & $0.592^{+0.011}_{-0.010}$ \\
NGC 6352 & $0.65 \pm 0.02$ & $4.56$  & $33.58$ & $12.70$ &  $0.474\pm0.035 $ & $0.51$ &  $0.475^{+0.028}_{-0.029}$ & $0.476^{+0.027}_{-0.024}$ \\
\rowcolor{Gray}
NGC 6362 & $1.27 \pm 0.03$ & $7.23$  & $51.84$ & $13.60$ &  $0.574\pm0.035 $ & $0.47$ &  $0.583^{+0.041}_{-0.039}$ & $0.585^{+0.038}_{-0.033}$ \\
NGC 6366 & $0.38 \pm 0.02$ & $5.56$  & $34.82$ & $13.30$ &  $0.418\pm0.045 $ & $0.64$ &  $0.355^{+0.027}_{-0.288}$ & $0.359^{+0.025}_{-0.022}$ \\
\rowcolor{Gray}
NGC 6388 & $12.50 \pm 0.11$ & $4.34$  & $100.55$ & $11.70$ &  $0.245\pm0.010 $ & $0.24$ &  $0.553^{+0.124}_{-0.139}$ & $0.726^{+0.009}_{-0.008}$ \\
NGC 6397 & $0.97 \pm 0.01$ & $3.90$  & $52.44$ & $13.40$ &  $0.345\pm0.036 $ & $0.40$ &  $0.466^{+0.023}_{-0.029}$ & $0.471^{+0.021}_{-0.019}$ \\
\rowcolor{Gray}
NGC 6496 & $0.69 \pm 0.07$ & $6.42$  & $28.78$ & $12.40$ &  $0.674\pm0.035 $ & $0.57$ &  $0.619^{+0.068}_{-0.059}$ & $0.619^{+0.068}_{-0.060}$ \\
NGC 6535 & $0.22 \pm 0.04$ & $3.65$  & $25.75$ & $10.50$ &  $0.536\pm0.081 $ & $0.65$ &  $0.427^{+0.084}_{-0.064}$ & $0.424^{+0.085}_{-0.063}$ \\
\rowcolor{Gray}
NGC 6541 & $2.93 \pm 0.09$ & $4.34$  & $38.94$ & $12.90$ &  $0.396\pm0.020 $ & $0.24$ &  $0.636^{+0.020}_{-0.017}$ & $0.638^{+0.019}_{-0.018}$ \\
NGC 6584 & $1.02 \pm 0.17$ & $5.37$  & $55.84$ & $11.30$ &  $0.451\pm0.026 $ & $0.45$ &  $0.498^{+0.049}_{-0.048}$ & $0.498^{+0.050}_{-0.050}$ \\
\rowcolor{Gray}
NGC 6624 & $1.56 \pm 0.04$ & $3.69$  & $19.45$ & $12.50$ &  $0.279\pm0.020 $ & $0.28$ &  $0.263^{+0.232}_{-0.131}$ & $0.499^{+0.012}_{-0.012}$ \\
NGC 6637 & $1.55 \pm 0.18$ & $3.69$  & $25.34$ & $13.10$ &  $0.425\pm0.017 $ & $0.28$ &  $0.591^{+0.036}_{-0.037}$ & $0.590^{+0.038}_{-0.036}$ \\
\rowcolor{Gray}
NGC 6652 & $0.48 \pm 0.07$ & $1.96$  & $18.84$ & $12.00$ &  $0.344\pm0.026 $ & $0.40$ &  $0.470^{+0.037}_{-0.032}$ & $0.467^{+0.038}_{-0.037}$ \\
NGC 6656 & $4.76 \pm 0.05$ & $5.29$  & $77.20$ & $12.70$ &  $0.274\pm0.020 $ & $0.25$ &  $0.376^{+0.179}_{-0.152}$ & $0.591^{+0.013}_{-0.013}$ \\
\rowcolor{Gray}
NGC 6681 & $1.16 \pm 0.02$ & $2.89$  & $27.72$ & $12.80$ &  $0.234\pm0.019 $ & $0.29$ &  $0.282^{+0.141}_{-0.184}$ & $0.471^{+0.009}_{-0.009}$ \\
NGC 6715 & $17.80 \pm 0.30$ & $5.20$  & $283.30$ & $10.80$ &  $0.267\pm0.012 $ & $0.24$ &  $0.742^{+0.031}_{-0.145}$ & $0.767^{+0.012}_{-0.011}$ \\
\rowcolor{Gray}
NGC 6717 & $0.36 \pm 0.08$ & $4.23$  & $19.80$ & $13.20$ &  $0.637\pm0.039 $ & $0.60$ &  $0.539^{+0.088}_{-0.080}$ & $0.541^{+0.089}_{-0.079}$ \\
NGC 6723 & $1.77 \pm 0.11$ & $5.06$  & $37.16$ & $13.10$ &  $0.363\pm0.017 $ & $0.33$ &  $0.515^{+0.023}_{-0.339}$ & $0.522^{+0.020}_{-0.019}$ \\
\rowcolor{Gray}
NGC 6752 & $2.76 \pm 0.04$ & $5.27$  & $67.43$ & $13.80$ &  $0.294\pm0.023 $ & $0.24$ &  $0.387^{+0.150}_{-0.207}$ & $0.534^{+0.014}_{-0.013}$ \\
NGC 6779 & $1.86 \pm 0.18$ & $4.51$  & $96.87$ & $13.70$ &  $0.469\pm0.041 $ & $0.29$ &  $0.629^{+0.051}_{-0.046}$ & $0.629^{+0.050}_{-0.045}$ \\
\rowcolor{Gray}
NGC 6809 & $1.93 \pm 0.08$ & $6.95$  & $51.52$ & $13.80$ &  $0.311\pm0.029 $ & $0.38$ &  $0.454^{+0.025}_{-0.299}$ & $0.465^{+0.019}_{-0.018}$ \\
NGC 6838 & $0.66 \pm 0.03$ & $6.57$  & $43.17$ & $12.70$ &  $0.622\pm0.038 $ & $0.58$ &  $0.540^{+0.046}_{-0.037}$ & $0.543^{+0.047}_{-0.040}$ \\
\rowcolor{Gray}
NGC 6934 & $1.36 \pm 0.19$ & $5.16$  & $86.40$ & $11.10$ &  $0.326\pm0.020 $ & $0.39$ &  $0.470^{+0.033}_{-0.029}$ & $0.466^{+0.036}_{-0.036}$ \\
NGC 6981 & $0.69 \pm 0.12$ & $5.96$  & $49.82$ & $10.90$ &  $0.542\pm0.027 $ & $0.55$ &  $0.495^{+0.057}_{-0.054}$ & $0.494^{+0.057}_{-0.055}$ \\
\rowcolor{Gray}
NGC 7078 & $6.33 \pm 0.07$ & $4.30$  & $140.75$ & $13.60$ &  $0.399\pm0.019 $ & $0.25$ &  $0.763^{+0.020}_{-0.019}$ & $0.763^{+0.020}_{-0.019}$ \\
NGC 7089 & $6.20 \pm 0.11$ & $4.77$  & $111.48$ & $11.80$ &  $0.224\pm0.014 $ & $0.25$ &  $0.199^{+0.245}_{-0.083}$ & $0.607^{+0.010}_{-0.009}$ \\
  \end{tabular}
  \caption{Properties of globular clusters as listed by \citet{Hilker20} and ages listed by \citep{Krause16}, with empirical unpolluted star fractions as listed by \citet{Milone17}. The last three columns are the fraction of these pristine stars in the model presented in this work assuming a total depletion factor $f_\mathrm{dep}=M_\mathrm{c}/M_\mathrm{c,0} = 0.5$, the $f_\mathrm{dep}$ inferred from an MCMC exploration with our model, and $f_\mathrm{dep,max}$ that is the result of the same MCMC exploration but assuming only unpolluted stars are removed from the cluster.  }
  \label{tab:1}
\end{table*}

\subsection{Abundance distribution synthesis}
\label{sec:compsynth}
\subsubsection{Synthesis procedure}

{In this section, we consider the elemental abundance variations within individual globular clusters. We first need to estimate the composition of the companions that form via accretion of the polluted ISM. We will assume that accretion is dominated by sweep-up by a long-lived disc (see discussion in Section~\ref{sec:origins}). In order to estimate the evolution of the ejecta abundances, we follow an approach similar to that of \citet{DErcole10}. As in Section~\ref{sec:rapid_acc}, we consider the quantity of mass added to the gas reservoir due to the evolution of massive stars. We consider the evolution of equation~\ref{eq:Mdot_gas} with the star formation rate:}
\begin{equation}
    \dot{M}_\mathrm{SF} = M_\mathrm{c,0} \cdot \frac{\epsilon_\mathrm{SF}}{\tau_\mathrm{ff}},
\end{equation}{for free-fall time $\tau_\mathrm{ff}$. The star formation efficiency per free-fall time $\epsilon_\mathrm{SF}$ is a free parameter. We will always assume $\epsilon_\mathrm{SF}=0$ initially, but `switch-on' star formation at a time $t_\mathrm{SF}$ defined as the MS lifetime for a star of mass $m_\mathrm{SF}$. We then draw new stars at times obtained from a probability distribution function proportional to $\dot{M}_\mathrm{SF}$, with a composition of pure ejecta material defined by the composition of the ISM at the time they form. }

{In order to simulate the evolution of the ISM composition, we keep track of the individual elements:}
\begin{equation}
\dot{M}_{\mathrm{el}, k} =  \alpha_{k}(t) \cdot \dot{M}_\mathrm{eject} - \beta_k \left( f_\mathrm{lm} M_\mathrm{c,0} \langle \dot{q}_\mathrm{sweep} \rangle + \dot{M}_\mathrm{SF} \right),
\end{equation}{where $k$ denotes an index referring to a specific chemical element, $\beta_k$ is the current mass fraction of the element $k$ in the reservoir, and $\alpha_{k}(t)$ is the mass fraction of element $k$ being ejected by the first generation stars. We estimate the specific yield of each element by interpolating between the model ejecta abundances of AGB ejecta \citep[as for][]{DErcole10}. For stars of mass $m_*>9\,M_\odot$, we use the massive binary ejecta abundances found by \citet{deMink09}. }

The cluster properties are taken from Table~\ref{tab:1}, while we assume that the initial cluster mass $M_\mathrm{c,0}= 5M_\mathrm{c}$, where $M_\mathrm{c}$ is the present day mass. The scale radius in a Plummer sphere where $a_\mathrm{stars} = R_\mathrm{hm}/1.3$ is fixed and we assume the gas is uniformly distributed with $R_\mathrm{gas}=a_\mathrm{stars}$. This also determines the central velocity dispersion. We then draw a sample of $400$ companions that share the abundances of the well-mixed ejecta at the time of their formation $t = t_\mathrm{form}$. The formation time is drawn from a probability density function which is proportional to $\langle \dot{q}(t)\rangle$, the mean rate of change of the companion mass ratio. {We initially fix the mixing fraction $f_\mathrm{mix}=0.7$ when modifying the primary star abundance in order to compare to observations, varying only the companion mass-ratio $q$. However, as per the discussion in Section~\ref{sec:outlook_abundances}, the required $q$ may be significantly smaller if $f_\mathrm{mix}$ is smaller by a factor of a few. We demonstrate how this works by two different approaches to modelling M54 (Section~\ref{sec:M54_popsynth}) and NGC 2808 (Section~\ref{sec:NGC2808_popsynth}). In the former, we assume a broad range of $q$, while in the latter we assume a narrow range of $q$ and vary $f_\mathrm{mix}$. }

\begin{figure}
    \centering
    \includegraphics[width=\columnwidth]{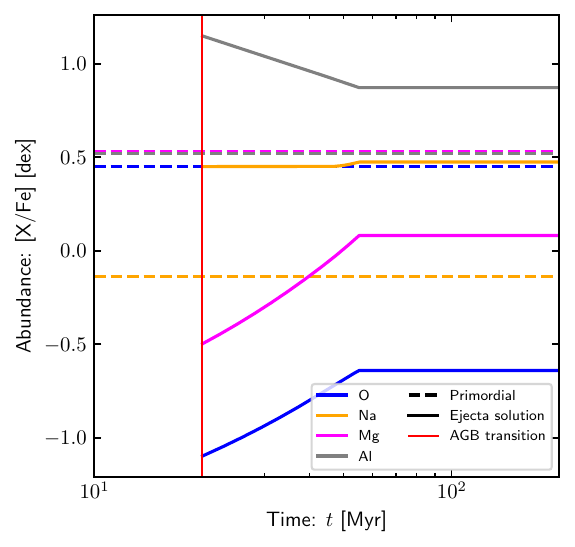}
    \caption{The abundances of elements in the gas reservoir of our model for 47 Tuc (solid lines) compared to the pristine abundances (horizontal dashed lines) as a function of time. Up to $20$~Myr we assume gas is expelled from the cluster. Afterwards we interpolate over the grid of ejecta used by \citet{DErcole10} for AGB stars. All elements are assumed to be instantaneously mixed with the existing gas reservoir at each time. The minimum stellar mass that contributes is $m_\mathrm{min}=6\, M_\odot$. After a star of $m_\mathrm{min}$ reaches the end of the MS abundances remain fixed.   }
    \label{fig:47Tuc_ejectabund}
\end{figure}

\begin{figure}
    \centering
    \includegraphics[width=\columnwidth]{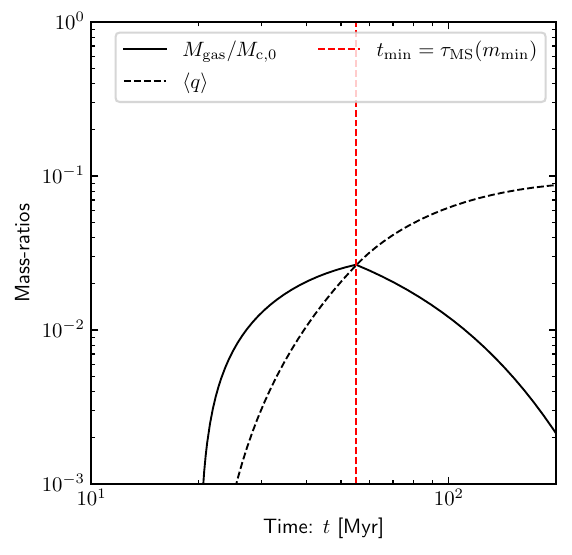}
    \caption{Total mass of ejecta in our model for 47 Tuc. The solid line is the total mass of ejecta material in the gas reservoir, the dashed line represents the average mass-ratio of companions. The dashed red vertical line shows when the ejecta contributions to the medium are `switched-off'. }
    \label{fig:47Tuc_mtotevol}
\end{figure}

{We do not produce a full model-fitting procedure. This is because, as discussed in Section~\ref{sec:abund_caveats}, there are numerous uncertainties and caveats in the model. This renders any outcomes of a fitting exercise difficult to interpret. However, we can vary some choice parameters in order to try to qualitatively fit some observed abundance distributions. In the following, we compare the outcome of our simple synthesis model to observed abundances in 47 Tuc, M54 and NGC 2808. For all these clusters we assume that a fraction $f_\mathrm{I}=0.25$ of the present day population had companions that underwent ionisation rather than accretion, consistent with our model predictions.}

\subsubsection{Population synthesis in 47 Tuc}

\label{sec:47Tuc_popsynth}
\begin{figure}
    \centering
   \subfloat[\label{subfig:47Tuc_ONa}]{\includegraphics[width=\columnwidth]{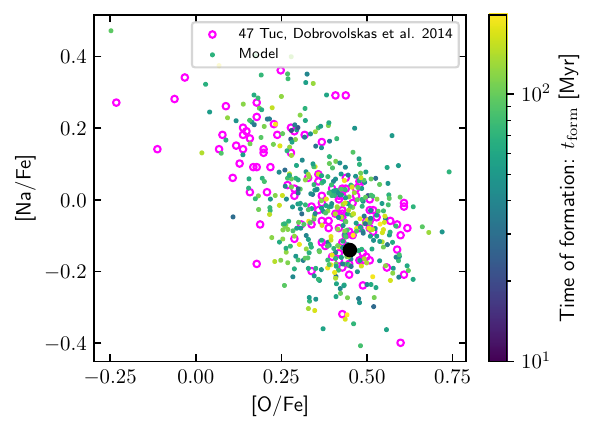}}\\
   \subfloat[\label{subfig:47Tuc_MgAl}]{\includegraphics[width=\columnwidth]{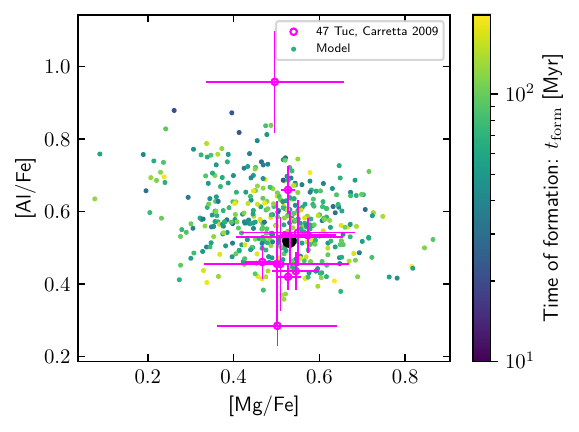}}
    \caption{Synthesised chemical enrichment of 47 Tuc stars according to our simple model as described in Section~\ref{sec:compsynth}. We show the assumed primordial abundances as black circles. Magenta circles show the abundances inferred by \citet{Dobrovolskas14} for Na and O (Figure~\ref{subfig:47Tuc_ONa}) and \citet{Carretta09b}
    for Mg and Al (Figure~\ref{subfig:47Tuc_MgAl}). A sample of $400$ Monte Carlo drawings from our model are shown as small circles coloured by the time at which the companion formed from the ejecta of massive binaries of AGB ejecta. Note that although all stars are assumed to form a companion, $25$~percent of these companions are ionised rather than accreted. The Na and O errors are not listed by \citet{Dobrovolskas14}, so we have assumed observational errors of $0.1$~dex in order to replicate observational scatter in our model sample. For the Al and Mg distributions, we adopt the median errors from the \citet{Carretta09b} sample. }
    \label{fig:47Tuc_synth}
\end{figure}

\begin{figure}
    \centering
    \includegraphics[width=\columnwidth]{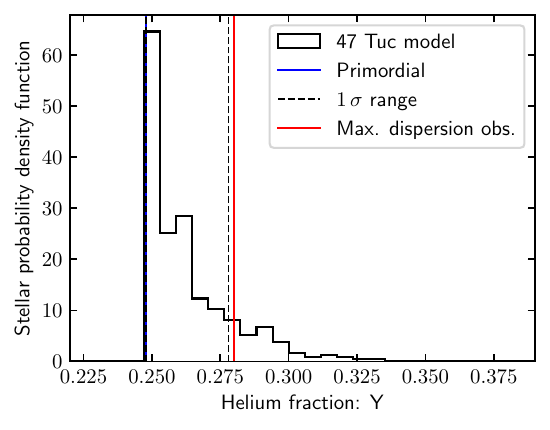}
    \caption{He distribution obtained from our model for the pollution in 47 Tuc (as in Figure~\ref{fig:47Tuc_synth}). We show the assumed primordial $Y=0.248$ as a vertical solid blue line and maximal dispersion $\Delta Y=0.03$ as a vertical red line inferred from constraints in the spread in the horizontal branch stars \citep{diCriscienzo10, Gratton13}. The range of the $1\,\sigma$ dispersion in the model is shown by vertical dashed lines. }
    \label{fig:He_47Tuc}
\end{figure}

{In 47 Tuc, we will find that we do not need to appeal to residual star formation from pure pollutant in order to reproduce the observed abundance variations. Nor do we require the contribution of both massive binaries and AGB ejecta. We therefore set $\epsilon_\mathrm{SF}=0$ throughout, and only include ejecta after $20$~Myr, the assumed MS lifetime of an AGB star of mass $m_\mathrm{max} = 9 \,M_\odot$. The minimum mass for the AGB stars contributing to the ejecta is $m_\mathrm{min}=6\,M_\odot$. The resultant abundance variations of the ISM and the total mass in the reservoir and companions are shown in Figures~\ref{fig:47Tuc_ejectabund} and~\ref{fig:47Tuc_mtotevol} respectively. In the latter, we see that the total gas reservoir never exceeds a few percent of the total initial cluster mass, while the average mass ratio of the companions eventually reaches $q\approx 0.1$.}

In order to estimate the abundance variations expected from our model, we draw 400 companions with a mass ratio $q$ from a log-normal distribution with median mass ratio $q_\mathrm{1/2}=0.1$ and $0.5$~dex dispersion. We assume pristine abundances $\rm{[O/Fe]}= 0.45$, $\rm{Na/Fe]}=-0.14$, $\rm{[Al/Fe]}=0.52$ and $\rm{[Mg/Fe]}=0.52$, and that a merger occurs for $75$~percent of the stars. We then compare to the abundances of O and Na inferred by \citet{Dobrovolskas14} and Mg and Al by \citet{Carretta09b}. We assume observational errors of $0.1$~dex for O and Na abundances \citep[abudance errors are not listed for these elements by][]{Dobrovolskas14}, and for Al and Mg abundances we take the median errors from \citet{Carretta09b}. We then offset the abundances by amount drawn from a Gaussian distribution with standard deviation determined by this observational error.

The companion accretion model is somewhat successful in reproducing the abundance variations in 47 Tuc, as shown in Figure~\ref{fig:47Tuc_synth}. The outcome is in reasonable agreement to the observed distributions, both for O and Na (Figure~\ref{subfig:47Tuc_ONa}) and Mg and Al (Figure~\ref{subfig:47Tuc_MgAl}). Some discrepancies between the observed abundances and our model are clear. {The synthetic O abundances exhibit a dispersion of $\sigma_\mathrm{mod} = 0.13$~dex, only slighly smaller than $\sigma_\mathrm{obs} =0.17$~dex in the \citet{Dobrovolskas14} sample. For Na, the observed and model dispersions agree at $0.15$~dex. The scatter in the Mg abundance in the model $\sigma_\mathrm{mod} = 0.12$~dex exceeds the observed dispersion $\sigma_\mathrm{obs} = 0.03$, however this is due to the quoted observational uncertainties rather than the physical model. The Al dispersion in the model ($\sigma_\mathrm{mod} = 0.1$~dex) is slightly lower than the observed ($\sigma_\mathrm{obs} = 0.16$~dex), with a single high Al data point that could be an outlier in this case.}

We also show the abundances of He for our model for 47 Tuc in Figure~\ref{fig:He_47Tuc}. The maximum dispersion in the He abundances inferred from horizontal branch stars is $\Delta Y \sim 0.03$ \citep{diCriscienzo10, Gratton13}. As pointed out by \citet{Bastian15b}, the largest O depletion values are inconsistent with this change in He abundances. Indeed, in Figure~\ref{fig:He_47Tuc} we see a tail of stars with larger He abundance variations. However, the dispersion in the model sample actually remains consistent with the small variation inferred from observations. {We interpret this success cautiously, since some other elements have a somewhat smaller dispersion than the observed abundances.} However, the upper limit in the He abundance dispersion of $0.03$ is only in slight tension with the \citet{DErcole10} ejecta abundances.

\subsubsection{Population synthesis in M54}
\label{sec:M54_popsynth}

\begin{figure}
    \centering
    \includegraphics[width=\columnwidth]{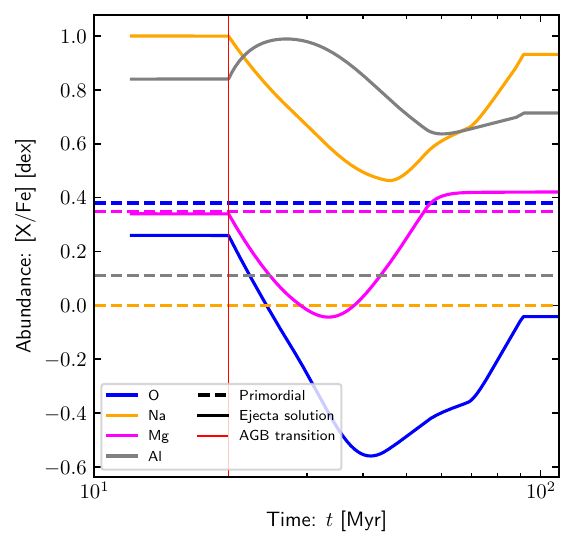}
    \caption{The abundances of elements in the gas reservoir of our model for M54. Line 
 colors and styles have similar meanings as in Figure~\ref{fig:47Tuc_ejectabund}.}
    \label{fig:M54_ejectabund}
\end{figure}

\begin{figure}
    \centering
    \includegraphics[width=\columnwidth]{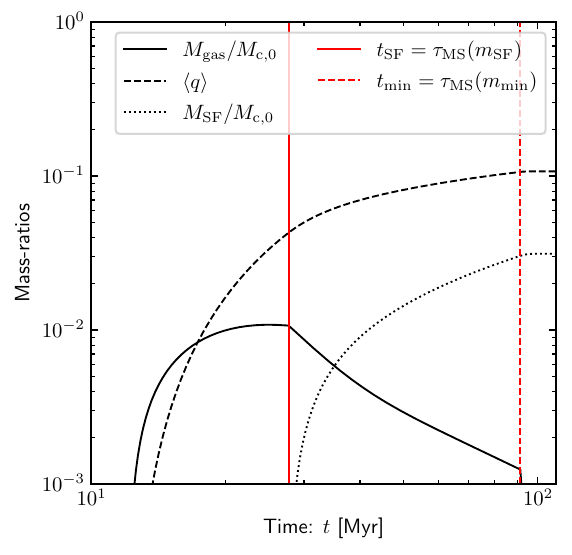}
    \caption{The relative mass in the gas reservoir (solid line) in second generation star formation (dotted line) and the average mass ratio (dashed) line in the model for M54. The solid red vertical line marks the time where star formation is switched on (stars of mass $m_\mathrm{SF}=8\, M_\odot$ reach the end of the MS), while the dashed red line marks the time where ejecta pollution of the environment is switched off (stars of mass $m_\mathrm{min}=5 \, M_\odot$ reach the end of the MS). We assume a star formation efficiency constant $\epsilon_\mathrm{SF} = 4\cdot 10^{-4}$. }
    \label{fig:M54_masses}
\end{figure}
\begin{figure*}
    \centering
    \subfloat[\label{subfig:M54_tAGB_ONa}]{\includegraphics[width=0.5\textwidth]{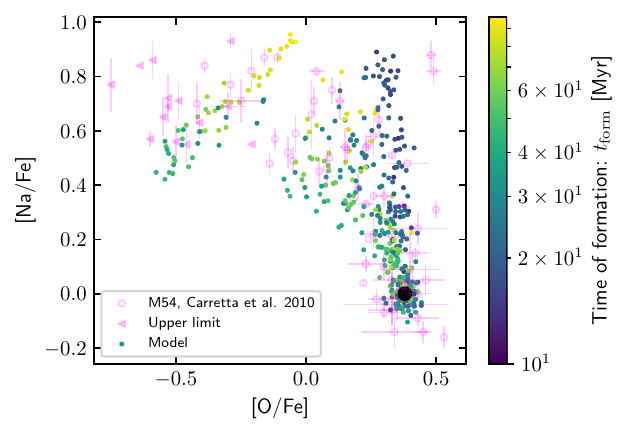}}
    \subfloat[\label{subfig:M54_tAGB_MgAl}]{\includegraphics[width=0.5\textwidth]{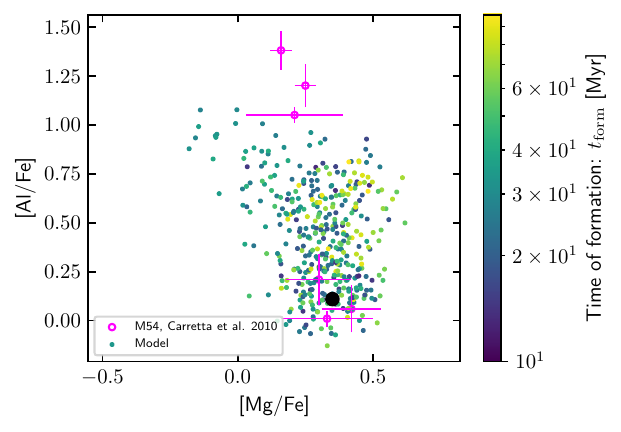}}
    \caption{As in Figures~\ref{fig:47Tuc_synth}, but for a model of M54 as described in the text. Here the observational data points are from \citet{Carretta10}. \label{fig:M54_popsynth} }
\end{figure*}

{We also apply our simple model to the the RGB star abundances in the older, more massive cluster M54 \citep{Carretta10}. In this case, as discussed in Section~\ref{sec:outlook_abundances}, we must appeal to residual star formation, and possibly some contribution from massive binaries to the available mass budget in order to produce the observed abundances. We here adopt $m_\mathrm{max} = 11\, M_\odot$, $m_\mathrm{min}=4.9 \, M_\odot$ and $m_\mathrm{SF} = 7.9\,M_\odot$, with a star formation efficiency per free-fall time $\epsilon_\mathrm{SF} = 4\times 10^{-4}$. We assume pristine abundances $\rm{[O/Fe]}= 0.38$, $\rm{Na/Fe]}=-0.06$, $\rm{[Al/Fe]}=0.11$ and $\rm{[Mg/Fe]}=0.35$. }

{The abundance evolution of the model ISM is shown in Figure~\ref{fig:M54_ejectabund}, while the mass evolution of the components of the system is shown in Figure~\ref{fig:M54_masses}. The latter indicates that the final $\langle q \rangle$ slightly exceeds $0.1$ and the total mass of stars formed from pollutant, $M_\mathrm{SF}$, is a few percent of the initial total cluster mass $M_\mathrm{c,0}$. If these stars are all low mass and do not get dynamically ejected from the cluster over time (due to formation in the core), then the fraction of them at the present day would be:}
\begin{equation}
    f_\mathrm{SF} = \frac{M_\mathrm{SF}}{M_\mathrm{SF} + f_\mathrm{dep} \cdot f_\mathrm{lm} \cdot M_\mathrm{c,0}}.
\end{equation}{We will here fix the depletion factor $f_\mathrm{dep}=0.5$ and the fraction of low mass stars $f_\mathrm{lm}=0.4$. In the case of our M54 model, this yields approximately $13$~percent of present day stars composed of pure ejecta.}

{Motivated by our results in Section~\ref{sec:outlook_abundances}, we adopt a log-uniform distribution between $\log q_\mathrm{min}=  -1.3$ and $\log q_\mathrm{max}=0.3$. This distribution of $q$ contains too many large $q$ values to be consistent with the mass budget. However, this may be mitigated in several ways. As discussed in Section~\ref{sec:outlook_abundances}, a somewhat reduced $f_\mathrm{mix}$ may explain the largest $q>1$, which we may not form frequently in the companion accretion model. When we model the population in NGC 2808 in Section~\ref{sec:NGC2808_popsynth}, we will use a more moderate range of $q$ and vary $f_\mathrm{mix}$ to demonstrate how this might work. In addition, the gas reservoir for forming companions may be  centrally concentrated, leading to higher $q$ companions in the central cluster stars that remain bound over Gyr timescales. Finally, the required mass-ratio for the highly polluted stars is strongly dependent on the model abundances of the pollutants; for example, slightly more Na and less O can significantly reduce the maximum required $q$.}

{We show the abundance variations in the model for M54 in Figure~\ref{fig:M54_popsynth}. We obtain {qualitative} agreement with the O and Na abundance distributions as shown in Figure~\ref{subfig:M54_tAGB_ONa}. In particular, the gap between the pure ejecta stars and those which have accreted a companion explains the substructure in the high Na, low O population. We obtain a model O dispersion $\sigma_\mathrm{mod} = 0.24$, moderately underestimating the observed $\sigma_\mathrm{obs}= 0.34$. For Na, we obtain $\sigma_\mathrm{mod} = 0.28$, close to the observed $\sigma_\mathrm{obs}=0.31$. In Figure~\ref{subfig:M54_tAGB_MgAl}, although the Mg distribution is reasonable ($\sigma_\mathrm{mode}= 0.14$, $\sigma_\mathrm{obs}= 0.17$), we find problems reproducing the Al distribution which is not (strongly) bimodal in our model.} 

{We conclude that aspects of the abundance variations in M54 can be well-produced by our model, while others suffer from the same problems as those of other self-enrichment scenarios. In general, our model can help to alleviate the mass budget problem by appealing to both massive binary and AGB ejecta and a variable $f_\mathrm{mix}$. If $f_\mathrm{mix} \sim 0.1{-}0.3$ for the subset of apparently high $q$ stars, then the observed abundance variations may be reproduced without requiring additional mass. However, even in this case, we are not be able to reproduce the bimodial Al distribution, or the most extreme O depletion in our model. This is not an issue of the mass budget, but must require additional physics -- for example, deviations from the theoretical model ejecta abundances.  }

\subsubsection{Population synthesis in NGC 2808}
\label{sec:NGC2808_popsynth}

\begin{figure}
    \centering
    \includegraphics[width=\columnwidth]{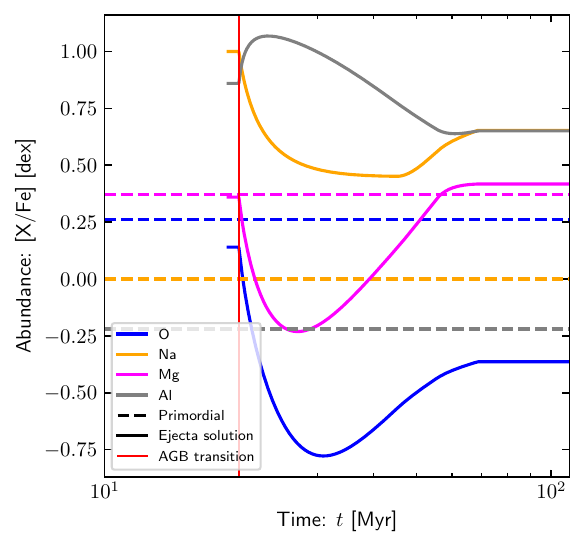}
    \caption{The abundances of elements in the gas reservoir of our model for NGC 2808. Line 
 colors and styles have similar meanings as in Figure~\ref{fig:47Tuc_ejectabund}.}
    \label{fig:NGC2808_ejectabund}
\end{figure}

\begin{figure}
    \centering
    \includegraphics[width=\columnwidth]{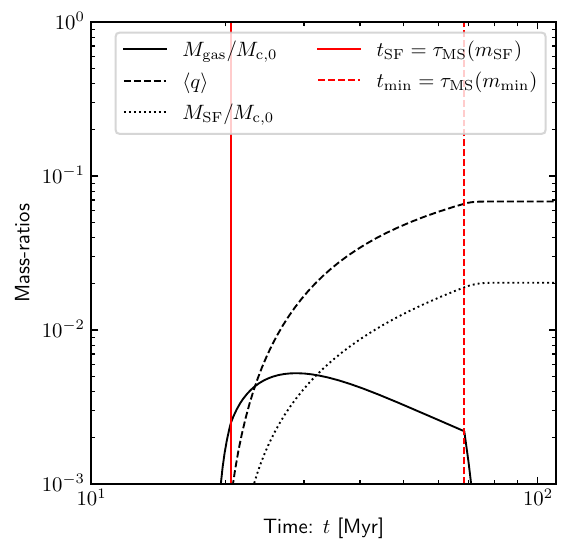}
    \caption{As in Figures~\ref{fig:47Tuc_mtotevol} and~\ref{fig:M54_masses} but for a model of NGC 2808. Free parameters are: $m_\mathrm{max} = 9.2 \, M_\odot$, $m_\mathrm{SF}=8.9\, M_\odot$,  $m_\mathrm{min}=5.5 \, M_\odot$, $\epsilon_\mathrm{SF} = 2.5\cdot 10^{-4}$.}
    \label{fig:NGC2808_masses}
\end{figure}

\begin{figure*}
    \centering
    \subfloat[\label{subfig:NGC2808_ONa}]{\includegraphics[width=0.5\textwidth]{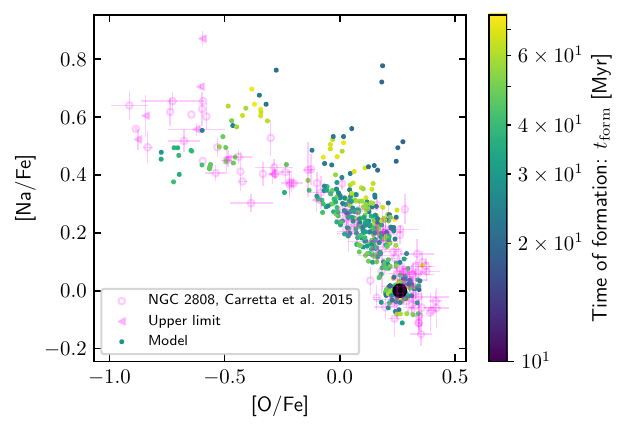}}
    \subfloat[\label{subfig:NGC2808_MgAl}]{\includegraphics[width=0.5\textwidth]{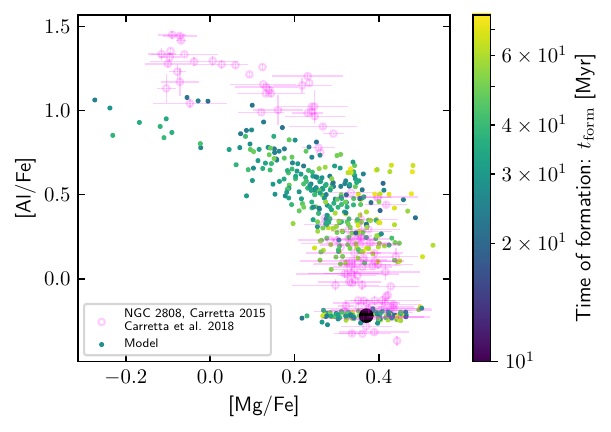}}
    \caption{As in Figures~\ref{fig:47Tuc_synth} and~\ref{fig:M54_popsynth}, but for a model of NGC 2808 as described in the text.\label{fig:NGC2808_fmixv}}
\end{figure*}

\begin{figure}
    \centering
    \includegraphics[width=\columnwidth]{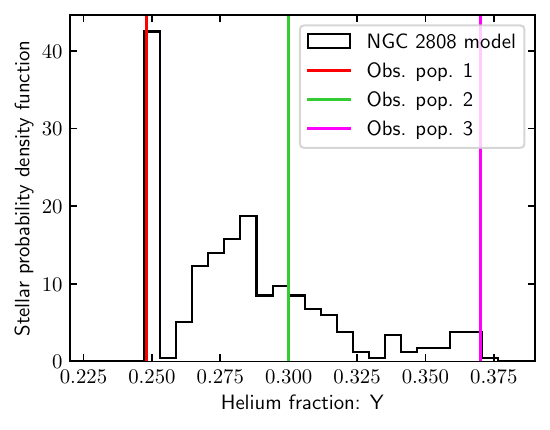}
    \caption{The helium abundance distribution obtained from our model for NGC 2808. The primordial abundance is assumed to be $Y=0.248$. We show the approximate abundances of the three populations identified by \citet{Piotto07} as coloured vertical lines. 
    \label{fig:He_NGC2808}}
\end{figure}

{Our final population synthesis effort is for NGC 2808, in which self-enrichment models have particular trouble to explain observed abundances \citep[see discussion by][]{DErcole10}. We compare our model with the O, Na and Mg abundances obtained by \citet{Carretta15} and Al abundances from UVES spectra obtained with FLAMES/GIRAFFE \citep{Carretta18}. The massive binary ejecta yield too much Na and O compared with the observed abundances in the RGB population of NGC 2808. We therefore consider a scenario where little of the massive binary ejecta is retained, adopting $m_\mathrm{max}=9.2\,M_\odot$. We also take $m_\mathrm{SF} = 8.9\,M_\odot$ with $\epsilon_\mathrm{SF} = 2.5\times 10^{-4}$ and $m_\mathrm{min} = 5.5\,M_\odot$. We assume pristine abundances $\rm{[O/Fe]}= 0.26$, $\rm{[Na/Fe]}= 0.0$, $\rm{[Mg/Fe]}=0.37$ and $\rm{[Al/Fe]}= -0.22$. The evolution of the elemental abundances is shown in Figure~\ref{fig:NGC2808_ejectabund}, while the mass evolution of the gas reservoir, average mass-ratio and residual star formation is shown in Figure~\ref{fig:NGC2808_masses}. In this case, due to the smaller mass range of the ejecta contributors, we obtain a slightly lower mass budget, with $\langle q \rangle \approx 0.07$. }

{We adopt a log-normal distribution in $q$ with a median mass-ratio $q_{1/2}=0.2$ that is moderately larger than that obtained directly from the integration, this time with $0.2$~dex dispersion. In order to obtain the strongly polluted stars, we will here adopt a range of $f_\mathrm{mix}$ rather than a broad range of $q$. In order to do this simply, we draw $f_\mathrm{mix}$ uniformly between $0.1$ and $0.7$, although highlight that this choice is arbitrary and intended only to illustrate the influence of $f_\mathrm{mix}$. }

{The results are shown in Figure~\ref{fig:NGC2808_fmixv}. As in M54, we can qualitatively reproduce some of the features seen in the abundance distributions. In both the O-Na case and Mg-Al cases, we reproduce the overall shape of the distribution. Here we also recover a multi-modal distribution in Al due to the clear distinction between the polluted and non-polluted stars. However, there is not the same large jump in Al abundances for low-Mg stars seen in the observed population in Figure~\ref{subfig:NGC2808_MgAl}. Possible origins for this discrepancy may be that mixing of the ejecta in the gas reservoir is inefficient, or if Al production is more efficient in massive binaries or AGB stars than predicted. Quantitatively, the model abundance dispersions are $\sigma_\mathrm{mod} = 0.18$, $0.22$, $0.12$ and $0.36$~dex for Na, O, Mg and Al respectively. Compared with the observed dispersions, $\sigma_\mathrm{obs} = 0.22$, $0.37$, $0.15$ and $0.57$~dex, Al is the only element that is more than $0.2$~dex underestimated in the model. The need for more Al in the ejecta is common to all three clusters we investigate in this work. In order to produce this Al, we may appeal to an additional progenitor, such as binary mergers for stars with masses lower than $9\,M_\odot$ \citep[see e.g.][and discussion Section~\ref{sec:abund_caveats}]{LongWang20}, to activate the appropriate reaction chains.}

{Finally, we show the He distribution in Figure~\ref{fig:He_NGC2808}. This is in broad agreement with the observed abundances in the triple main sequence in NGC 2808, which has discrete populations with inferred $Y\approx 0.248$, $0.3$ and $0.35-0.40$ \citep{Piotto07}. The exact relative sizes and distinctness of these three populations is dependent on the star formation efficiency, the companion mass-ratio distribution and mixing fractions.  }



\subsubsection{Caveats and the origins of the model short-comings}
\label{sec:abund_caveats}

{The very simple model for chemical enrichment comes with numerous caveats, which may help or hinder the reproduction of the observed abundances. Some such considerations are as follows:}
\begin{itemize}
     \item {AGB (and massive binary) ejecta abundances are model dependent \citep[e.g.][]{Ventura05a, Ventura05b, Karakas07}. For example, \citet{Ventura09} predict lower Mg and higher Al yields, which would help in the case of the comparisons to the M54 and NGC 2808 abundances. On the other hand, the hotter temperatures in massive stars or binary interactions may be needed to activate the MgAl chain \citep[for example, see][and references therein]{Prantzos06,Kobayashi20}. {Frequent early binary mergers, possibly driven by cluster dynamics  \citep{LongWang20}, may further contribute to the pollutants, increasing both Al content and mass budget from which to form companions}. Overall, the predictive power of AGB ejecta abundances is limited, and this problem is inherent in all self-enrichment models.}
    \item {The internal sources of pollution via the merging, rotational mixing and later dredge-up processes are not accounted for in the model. Any combination of these processes may further alter the abundances in the collision product from the pristine values, as supported by a number of observational constraints discussed in Section~\ref{sec:internal}. }
    \item {Further to the above, if we assume that in some cases a minority of stars can have massive polluted companions $q\sim 1$, then the internal burning products of both stars, not just the host, can contribute to surface abundance variations. }
\end{itemize}

\subsubsection{Outlook for producing abundance variations by companion accretion}
\label{sec:outlook_abund_2}
{We have demonstrated that under feasible assumptions, the companion accretion model can reproduce many aspects of the observed abundance distributions in specific stellar clusters. {However, in the cases of the massive and old clusters M54 and NGC 2808, we have needed to assume some moderate enhancement of the mass ratios of the companions, as well as variable mixing and residual star formation.} There remain uncertainties about the degree of feasible mixing and the maximum possible mass ratio of a companion, while Al abundances in particular requires appealing to uncertainties in AGB ejecta composition. A number of possible avenues exist for more detailed study of the evolution of the collision product post-accretion. Overall, we conclude that the abundances we obtain from our models are promising, but not yet conclusive.}

\subsection{Stellar property trends }

\subsubsection{Binarity correlations}

We expect the mechanism we have described to naturally give rise to the observed absence of enrichment among binary stars \citep[e.g.][]{DOrazi15}. This is because the triple system will preferentially eject the lower mass component once a high eccentricity is excited \citep[e.g.][]{Anosova86}, and hence lose the potentially polluting companion.

\subsubsection{Central concentration}

In the companion accretion model, we expect that the polluted companions are accreted most frequently at high velocity dispersions in the central regions of a cluster. This would be in line with several observed regions \citep[e.g.][]{Simioni16}. In general, we expect the dynamical mixing over time to dilute the polluted population, although since our model allows the pollution to occur much later than other models, this should be the optimum scenario for producing spatial segregation of the two populations. Clusters that do not show spatially segregated populations may still be consistent with dynamical mixing \citep[e.g.][]{Vanderbeke15}. {It is less obvious whether some circumstances may precipitate an outcome in some clusters where the polluted  
population is actually {\it less} centrally concentrated than the pristine stars \citep[e.g.][]{Larsen12}.} {The rate of companion capture is dependent on the local density and velocity dispersion of the stars and gas reservoir and the localised star formation rate.} Hence it is possible that in some birth environments the polluted material is accreted most efficiently outside of the high velocity central regions. However, the nature of this scenario is dependent on the (turbulent) properties of the wind ejecta, and we do not explore this further here.

\subsection{Summary and caveats}

\subsubsection{Successes of the companion accretion model}

We have presented a model for producing multiple stellar populations in globular clusters that hinges on the accretion of polluted sub-stellar companions due to eccentricity excitation by stellar encounters. This mechanism allows us to make quantitative predictions for the rate of enrichment, which agree with those observed. Our model offers the following successes, many of which are not achieved by existing models:
\begin{itemize}
    \item \textit{Presence of absence of multiple populations:} The rate of accretion of companions is a predictable consequence of averaged dynamical encounters. {In general our model can accurately predict which clusters exhibit evidence of multiple populations. In a small number of clusters there is some minor in tension between our model and the presence or absence of multiple populations. However, this is only in clusters where the sample sizes are small (Ruprecht 106) or where the populations are not unambiguously identified (NGC 2155).} 
    \item \textit{Age correlation:} The enrichment in the companion accretion model typically occurs on Gyr timescales, such that we expect an absence of polluted stars for young clusters that is consistent with observations \citep{Martocchia18a}.
    \item \textit{Scaling of pollution fraction with cluster properties:} Our model is in excellent agreement with the observed fractions of polluted populations, reproducing similar scalings with the cluster properties for an assumed depletion factor $f_\mathrm{dep}=0.5$. Conversely, the observed fractions can be reproduced by appealing to moderate depletion factors, in contrast to many multi-generational models \citep[e.g.][]{Conroy11}.  
    \item \textit{Binarity:} Our model predicts lower rates of polution among binary stars \citep[e.g.][]{DOrazi15}, since a lower mass companion in an unstable configuration would be dynamically ejected rather than accreted. 
     \item \textit{Discreteness:} {{We have shown in Section~\ref{sec:compsynth} that whether or not the companion merges with its host, and different ejecta material and mixing fractions can produce complexity in the elemental `chomosome' maps \citep{Milone17} and the multi-modal He distribution as inferred from the multiple main sequences in some clusters \citep[e.g.][]{Piotto07}.}}
\end{itemize}

{There are further possible advantages of our model that remain more speculative: }
\begin{itemize}

    \item {\textit{Dilution and mass budget:} In the scenario we explore in this work, the dilution of the heavier elements occurs within the pristine star itself, long after all the ejecta from massive and AGB stars in the cluster has been expelled or accreted. {We have shown that a mass ratio $q\sim 0.1$ of a companion composed of polluted material mixed through $f_\mathrm{mix}=70$~percent of the primary is sufficient to produce typical abundance variations in some clusters (47 Tuc is explored here). However, late collapse of the gas reservoir and variable mixing fractions are required to explain the most extreme abundance variations while maintaining a small $q$. In the old and massive clusters M54 and NGC 2808, similar abundance variations to those observed can be reproduced in this way, but we must also appeal to order unity enhancements in the typical mass-ratios of companions. These may be possible if the gaseous reservoir is centrally concentrated. {Alternatively additional physics may contribute to surface abundance variations, such as mixing internal burning products in the primary into the convective region as a result of the merger.}}}
    \item \textit{Dynamical removal of pristine stars:} {Two-body encounters that would ionise the companion from its host star may be correlated to those that impart large amounts of kinetic energy into the trajectory of a star. It is possible that this offers a mechanism for preferential depletion of the pristine population, on top of the central concentration of the stars that are expected to undergo collisions with their companion. However, this possibility must be investigated with detailed dynamical models. }
\end{itemize}

\subsubsection{Uncertainties and caveats}

While the accretion rate of companions at a given {stellar} density and velocity dispersion is well-quantified as a function of time in this work, the feasibility of a number of aspects of the model require more detailed examination in future. The most serious of these are:
\begin{itemize}
\item \textit{Formation of the sub-stellar companion(s):} {We have invoked a number of possible mechanisms to produce sub-stellar companions, but have particularly adopted the rate at which a long-lived, compact disc sweeps-up material \citep[cf.][]{Bastian13}. However, the gas capture scenarios we discuss require follow-up with hydrodynamics simulations. {This formation process may also be in competition with ram pressure stripping of the (structured) gas reservoir \citep{Chantereau20}.}}
\item \textit{Stellar mixing:} {The chemical signatures in our model are dependent on the degree of mixing. We have appealed to simulations of grazing, high mass ratio mergers and mixing performed by \citet{Lombardi02} and \citet{Cabezon22}. These simulations suggest that deep mixing through the majority of the star is possible, but depends on the nature of the collision. Further quantification of the degree and homogeneity of the rotational mixing post-merger, and the subsequent return to the MS are required to fully test the expected changes to the surface abundances.}
\item \textit{Lithium survival:} {Linked to the above, it is unclear whether we should expect lithium to survive after a merger with a low mass companion. The models of \citet{Lombardi02} and \citet{Cabezon22} show survival of lithium on the stellar surface during the merger. However, these models have not been performed for an initially Li depleted companion, while the long term survival is also dependent on subsequent mixing.}
\item \textit{Ejecta content:} We have assumed that it is possible to produce the required pollutants out of some combination of the massive and AGB ejecta. In general, the wind abundances, particularly for AGB stars, are highly uncertain \citep[e.g.][]{Ventura05a, Ventura05b, Renzini13}. This remains true for any scenario that invokes first generation stars within the cluster as a source for abundance variations. 
\item \textit{Initial eccentricity of the companion:} {We have assumed an initial eccentricity of the companion $e_0=0.5$, which the rate at which collisions occur in the cluster. We make this choice as the mean of a uniform eccentricity distribution, which appears to characterise binaries of separation $\sim 100$~au \citep{Hwang22}. Such binaries may have formed via a similar process of fragmentation of a disc. However, it is unclear if such a process is similar for the formation process invoked in the context of this work.} 
\end{itemize}

\subsubsection{Future tests of the companion accretion model}

{While the companion accretion model is feasibly consistent with current observation constraints, future empirical tests may falsify the companion accretion model:}
\begin{itemize}
    \item {\textit{Companions in stellar clusters:} By far the most direct test of the model is the search for $q\sim 0.1$ companions in surveys of massive clusters aged $\sim 1{-}4$~Gyr, before most have been ionised or accreted. The main problem with this is that long baseline measurements are required to detect companions at $\sim 5$~au separations in radial velocity surveys. Unambiguously identifying companions becomes particularly problematic for high orbital eccentricities in dense clusters, when dynamical interactions with neighbours may also accelerate the star. However, such an investment may have numerous benefits beyond a test of the companion accretion model.}
    \item {\textit{Chemistry of RGB `companions':} As discussed in the introduction, some RGB stars appear to host companions that may be undergoing disruption or exchange of material with their host \citep{Soszynski21}. These companions appear to also exist in globular clusters \citep{Percy21}. If these examples represent the leftover survivors of companion accretion, then studying these stars and their companions for abundance variations may offer a window into this process.}
    \item {\textit{Companion merger simulations:} Further testing of the companion accretion model may be based on predictions from detailed simulations for the mergers, including the post-merger evolution of the remnant. These simulations may be compared quantitatively with observational constraints from Hertzsprung-Russell contours, for example \citep[cf. the early interactions of massive close binaries by][]{Wang20}.}
\end{itemize}

\section{Conclusions}
\label{sec:concs}
In this work, we have explored a novel mechanism for producing multiple stellar populations in globular clusters: `the companion accretion model'. In this scenario, the enriched wind ejecta from massive and AGB stars forms a bound gaseous reservoir, which remains unable to {form stars efficiently} for $\sim 100$~Myr timescales \citep[e.g.][]{Conroy11}, which may depend on the density of the gas \citep{Yaghoobi22}. {Stars moving through this medium may accrete gas from this reservoir by a combination of disc-sweeping, tidal cloud capture and Bondi-Hoyle-Lyttleton accretion.} This material may subsequently cool and collapse to produce a sub-stellar companion(s) of mass ratio $q\sim 0.1$. Over Gyr timescales, companions undergo eccentricity excitations due to dynamical encounters that can result in a grazing collision with the host star \citep{Kaib14, Hamers17, Winter22c}. {Such a collision may inject the pollutant in the secondary \citep{Lombardi02, Cabezon22}, which can broadly produce typical abundance variations in globular clusters. Extreme abundances may be produced by the leftover pollutant that remains bound and is allowed to cool and collapse to form a small second population from pure pollutant in some clusters.}

{The fraction of pristine stars in this model can be calculated, in a given environment by considering the ratio in which stellar encounters induce ionisations compared with collisions, also factoring in the effect of dynamical depletion of ionised (hence pristine) systems from the cluster core. We find that the predicted fraction of pristine stars as a function of environment is in good agreement with observations, requiring only moderate (factor two) depletion of the stellar mass at the present day.}

{This model has numerous benefits over existing models. It provides a natural way to produce ubiquitous multiple populations in dense and old clusters \citep[e.g.][]{Milone17} and a dearth of multiple populations in young clusters \citep[e.g.][]{Martocchia18a}. Companion accretion offers a comparatively simple dilution mechanism, without needing to retain primordial gas \citep[e.g.][]{DErcole11}. Most notably, the companion accretion model predicts a correlation of the enrichment fraction with cluster mass, or more precisely the internal velocity dispersion, since ionisation of the companions is less efficient for high velocity encounters.}

{There remain some important assumptions and possible sources of tension between our model and the observational constraints. For example, resolving the mass budget problem requires a variable mixing fraction, residual star formation and moderate enhancements to the average $q$ for stars that survive in a globular cluster to the present day. Particular future attention may be given to an exploration of the initial accretion and gravitational collapse of the companion from the polluted ejecta in the turbulent medium \citep[e.g.][]{Kuffmeier20}, the mixing of pollutant in the primary \citep[e.g.][]{Lombardi02, Cabezon22}, the uncertainties in the yield of elements in massive star and AGB ejecta \citep[e.g.][]{Ventura05a, Ventura05b, Renzini13}, and the stripping of the structured gas reservoir \citep{Chantereau20}. Each constitutes a complex process, for which the conclusions of this manuscript offer motivation for further consideration.}

\section*{Data availability}

The data used in this manuscript is all publicly available. Any scripts used for making figures shown are available from the corresponding author upon reasonable request.  

\section*{Acknowledgements}
{We sincerely thank the anonymous referee for an extremely useful referee report, which had a substantial impact on aspects of the model presented in this work as well the clarity of the manuscript.} AJW thanks the Institute of Astronomy, Cambridge for the funding for the visit in which this manuscript was written. We further thank Chris Tout and Arnab Sarkar for useful discussion regarding stellar mixing. In addition, AJW thanks Kees Dullemond, Maria Bergemann and Ivan Cabrera-Ziri for discussions on aspects of the topics covered in this manuscript. This project has received funding from the European Research Council (ERC) under the European Union’s Horizon 2020 research and innovation programme (PROTOPLANETS, grant agreement No. 101002188).




\bibliographystyle{mnras}
\bibliography{bdbib} 
\appendix

\section{Pristine star dynamical depletion}
\label{app:reldep}

We wish to quantify the relative depletion of the pristine and polluted population. Unfortunately, without more information, equation~\ref{eq:fdep} is the only constraint available to us linking $f_\mathrm{I,dep}$ and $f_\mathrm{II,dep}$, so that to go further we must make a choice on the form of these functions. 

A minimum for $f_\mathrm{I,dep}$ can be found by enforcing $f_\mathrm{II,dep}\leq 1$, such that:
\begin{equation}
\label{eq:fI_min}
    f_\mathrm{I,dep} \geq f_\mathrm{I,min}^{(1)} = f_\mathrm{dep} \cdot  \frac{1 - P_\mathrm{coll}/f_\mathrm{dep}}{1-P_\mathrm{coll}} .
\end{equation}When the fraction of unpolluted stars $ (1-P_\mathrm{coll})$ is large, adopting $ f_\mathrm{I,dep} = f_\mathrm{I,min}$ (i.e. all lost stars are not polluted) makes sense, since the vast majority of lost stars should anyway be unpolluted. This is because polluted stars will preferentially occupy the core region of the cluster, where they are most likely to remain bound, as discussed in Section~\ref{sec:depletion}. 

As $P_\mathrm{coll}$ grows, possibly exceeding $f_\mathrm{dep}$ to yield $f_\mathrm{I,dep}<0$, then this choice must eventually underestimate the loss of the polluted population II stars. Thus, we impose another minimum:
\begin{equation}
\label{eq:fI_min2}
    f_\mathrm{I,dep} \geq f_\mathrm{I,min}^{(2)} =  f_\mathrm{dep} \cdot P_\mathrm{coll}.
\end{equation}We choose this minimum because it remains greater than zero for all $f_\mathrm{dep}>0$ and $P_\mathrm{coll}>0$, as well as scaling with, but is always smaller than, the overall $f_\mathrm{dep}$ (as we expect). This minimum also scales with the overall fraction of polluted stars, which means that as the enrichment fraction grows the unpolluted stars become comparatively less likely to be ejected. This is what we expect, given that a large $P_\mathrm{coll}$ should result in less central concentration of the polluted stars.

In practice, we take the maximal of these two minima as $f_\mathrm{I,dep}$, from which we can obtain the corresponding $f_\mathrm{II,dep}$ for any given $f_\mathrm{dep}$ and $P_\mathrm{coll}$. The scaling of many of our results do not depend strongly on this choice, however we highlight that the form of $f_\mathrm{I,dep}$ and $f_\mathrm{II,dep}$ should be the study of further work. They are uncertain here as they are for other models for multiple populations.

\bsp	
\label{lastpage}
\end{document}